\documentclass[useAMS,usenatbib,usedcolumn]{mn2e}

\usepackage{graphicx,color,txfonts,longtable} 
\usepackage[authoryear]{natbib}
\usepackage{multicol}

\newfont{\gwpfont}{cmssq8 scaled 1000}
\newcommand{\reflex}{{\gwpfont REFLEX}}
\newcommand{\rtwo}{{\gwpfont REFLEX II}}
\newcommand{\rosat}{ROSAT}

\title[Construction of REFLEX II superclusters]
{The extended ROSAT-ESO Flux Limited X-ray Galaxy Cluster Survey (REFLEX II) \\
III. Construction of the first flux-limited supercluster sample
\thanks{Based on observations at the European Southern Observatory La Silla, Chile}}
\author[G. Chon, H. B\"ohringer and N. Nowak]{Gayoung Chon$^{1}$\thanks{E-mail:
gchon@mpe.mpg.de}, Hans B\"ohringer$^{1}$ and Nina Nowak$^{2}$\\
$^{1}$Max-Planck-Institut f\"ur extraterrestrische Physik, D85748 Garching, 
Germany \\
$^{2}$Max-Planck-Institut f\"ur Physik, F\"ohringer Ring 6, D80805 M\"unchen, 
Germany}
\begin{document}

\date{Accepted 2012 December 7 Received 2012 December 6; in original form 2012 November 12}

\pagerange{\pageref{firstpage}--\pageref{lastpage}} \pubyear{2012}

\maketitle

\label{firstpage}

\begin{abstract}

We present the first supercluster catalogue constructed with the extended 
ROSAT-ESO Flux Limited X-ray Galaxy Cluster survey (\rtwo) data, which 
comprises 919 X-ray selected galaxy clusters with a flux-limit of 
$1.8\times10^{-12}$~erg s$^{-1}$ cm$^{-2}$. Based on this cluster catalogue we 
construct a supercluster catalogue using a friends-of-friends algorithm with 
a linking length depending on the (local) cluster density, which thus varies 
with redshift. The resulting catalogue comprises 164 superclusters at 
redshift $z\leq0.4$. 

The choice of the linking length in the friends-of-friends method modifies 
the properties of the superclusters. We study the properties of different
catalogues such as the distributions of the redshift, extent and multiplicity 
by varying the choice of parameters. In addition to the supercluster 
catalogue for the entire \rtwo\ sample we compile a large volume-limited 
cluster sample from \rtwo\ with the redshift and luminosity constraints 
of $z\leq0.1$ and $L_X\geq5\times10^{43}$ erg/s. With this catalogue we 
construct a volume-limited sample of superclusters. This sample is 
built with a homogeneous linking length, hence selects effectively the same type of 
superclusters. By increasing the luminosity cut we can build a hierarchical 
tree structure of the volume-limited samples, where systems at the top of the
tree are only formed via the most luminous clusters. This allows us to test
if the same superclusters are found when only the most luminous clusters are
visible, comparable to the situation at higher redshift in the \rtwo\ 
sample. We find that the selection of superclusters is very robust, independent 
of the luminosity cut, and the contamination of spurious superclusters among
cluster pairs is expected to be small. 

Numerical simulations and observations of the substructure of clusters 
suggest that regions of high cluster number density provide an astrophysically 
different environment for galaxy clusters, where the mass function and X-ray
luminosity function are shifted to higher cut-off values and an increased
merger rate may also boost some of the cluster X-ray luminosities.
We therefore compare the X-ray luminosity function for the clusters
in superclusters with that for the field clusters with the flux- and volume-limited catalogues. The results mildly support the theoretical suggestion of a 
top-heavy X-ray luminosity function of galaxy clusters in regions of high 
cluster density.
\end{abstract}

\begin{keywords}
cosmology: large-scale structure of Universe -- Xrays:galaxies:clusters -- galaxies:clusters:general
\end{keywords}

\section{Introduction}

Superclusters are the largest, prominent density enhancements in the Universe. 
The first evidence of superclusters as agglomerations of rich clusters of 
galaxies was given by \citet{abell-61}. The existence of superclusters was 
confirmed by \citet{bogart-73}, \citet{hauser-73} and \citet{peebles-74}. 
Several supercluster catalogues based on samples of Abell/ACO clusters of 
galaxies followed, e.g. \citet{rood-76}, \citet{thuan-80}, \citet*{bahcall-84}, \citet*{batuski-85}, \citet{west-89}, \citet{zucca-93}, \citet*{kalinkov-95}, \citet{einasto-94,einasto-97,einasto-01,einasto-07}, and \citet{liiv-12}, among others. \defcitealias{einasto-01}{EETMA} \citet{einasto-01}, hereafter 
\citetalias{einasto-01} were the first using X-ray selected clusters
as well as Abell clusters.

Superclusters are generally defined as groups of two or more galaxy
clusters above a given spatial density enhancement~\citep{bahcall-88}. 
Their sizes vary between several tens of Mpc up to about $150$~$h^{-1}$~Mpc. 
As the time a cluster needs to cross a supercluster is larger than the age 
of the Universe superclusters cannot be regarded as relaxed systems. 
Their appearance is irregular, often flattened, elongated or filamentary 
and generally not spherically symmetric. This is a sign that they still 
reflect, to a large extent, the initial conditions set for the structure 
formation in the early Universe. They do not have a central concentration, 
and are without sharply defined boundaries.

Compared to this earlier work we have made progress in compiling 
statistically well-defined galaxy cluster catalogues. X-ray galaxy 
cluster surveys allow us to construct nearly mass-selected cluster catalogues 
with well understood selection functions. The purpose of this work is the 
construction of a supercluster catalogue on the basis of \rtwo\ -- a 
complete and homogeneous sample of X-ray selected galaxy clusters -- and a 
detailed analysis of its characteristics. X-ray selected galaxy clusters are 
very good tracers of the large-scale structure as their X-ray luminosity 
correlates well with their mass. So far numerous supercluster catalogues have 
been published based essentially on optically selected samples of galaxy 
clusters. Such samples suffer among other things from projection effects, 
whereas X-ray selected cluster samples provide a well-known selection function.
Here a novel method of compiling such a catalogue is presented which accounts 
for the specific properties of X-ray selected cluster samples. The resulting 
catalogue comprises 164 superclusters in the redshift range $z\leq0.4$ and is 
in good agreement with existing supercluster catalogues. 

A main goal of this work is the characterisation of superclusters as distinct
environments compared to the field, and the use of superclusters as special
laboratories making use of the particular environmental conditions. 
One application in this paper is the comparison of the X-ray luminosity 
function of galaxy clusters in superclusters and in the field. We expect
some difference for the following reason. According to Birkhoff's theorem
structure evolution in a supercluster region can be modelled in an equivalent
way to a Universe with a higher mean density than that of our Universe.
The major difference between these two environments is then a slower growth
of structure in the field compared to the denser regions of superclusters
in the recent past. Thus, as will be further explained in Sec. 7, we would 
expect a more top-heavy X-ray luminosity function in superclusters. We 
subject our supercluster sample to a test of this expectation.

This paper is organised as follows. After a brief description of the 
\rtwo\ catalogue in Sec. 2, we explain the construction of the 
supercluster catalogue in Sec. 3. In detail we also illustrate the role of the 
overdensity parameter, and give a justification for our parameter choices. 
Sec. 4 presents the \rtwo\ supercluster catalogue and gives an overview 
of the properties of the superclusters. A more detailed view on 
the catalogue is given in Sec. 5. We take a volume-limited sample from 
the \rtwo\ catalogue to construct a basis for the study of more 
astrophysical aspects of superclusters in Sec. 6 and study the supercluster 
selection by varying the overdensity we probe. In Sec. 7 we study the X-ray 
luminosity functions for the clusters, and compare them to the field clusters 
to look for differences due to environmental effects. We summarise and give 
an outlook of our future work in Sec. 8.

For the derivation of distance-dependent parameters, we use a flat 
$\Lambda$-cosmology with $\Omega_{\mathrm M}$=0.3, and 
$H_{\mathrm 0}=100~h$~km s$^{-1}$ Mpc$^{-1}$ with $h=0.7$. 
We note that the resulting supercluster catalogue does not depend
on the energy density parameters.

\section{REFLEX II cluster catalogue}

The \reflex\ (ROSAT-ESO Flux Limited X-ray) cluster survey is based
on the ROSAT All-Sky Survey (RASS;~\citet{truemper-93}). Our main 
goal with this project is to characterise the large-scale structure 
and to utilise them as astrophysical and cosmological probes. 
The extended \reflex\ cluster sample, \rtwo\, comprises 919 
clusters of galaxies with its last spectroscopic campaign completed 
in 2011~\citep{chon}. The clusters are located in the southern hemisphere 
below a declination of +2.5~degrees excluding the region around the 
galactic plane ($\pm$20$^\circ$) as well as the regions covered by the Large and the 
Small Magellanic Clouds. 
The total survey area is 4.24~sterad (33.75\% of the entire sky). 
The \rtwo\ clusters are selected with a flux-limit of $1.8\times10^{-12}$ 
erg s$^{-1}$ cm$^{-2}$ in the \rosat\ energy band (0.1--2.4~keV). 
\rtwo\ is a homogeneous sample with an estimated high completeness of 
more than 90\%. The highest cluster redshift is $z=0.539$, and roughly 
98\% of the clusters are at redshifts $z\leq0.4$.

The flux limit was imposed on a fiducial flux calculation assuming cluster 
parameters of 5 keV for the ICM temperature, a metallicity of 0.3 $Z_{\odot}$, 
a redshift of $z$=0, and an interstellar hydrogen column-density according to 
the 21cm measurements of~\cite{hi}. This fiducial flux was calculated 
independent of (prior to) any redshift information and is therefore somewhat 
analogous to an extinction-corrected magnitude limit without K-correction in 
optical astronomy. After measuring the redshift with our spectroscopic 
campaigns we then calculated the true fluxes and luminosities by taking an 
estimated ICM temperature from the X-ray luminosity-temperature scaling 
relation~\citep{pratt-09} and redshifted spectra into account.

The total count rates, from which fluxes and luminosities
were determined, were derived with the growth curve analysis (GCA)
as described in~\cite{boehringer-00}. The cluster candidates were 
compiled from a flux-limited sample of all sources in the RASS 
analysed by the GCA method, combining all information on the X-ray 
detection parameters, visual inspection of available digital
sky survey images, information in the NASA extragalactic
database\footnote{The NASA/IPAC Extragalactic Database (NED) 
is operated by the Jet Propulsion Laboratory, California 
Institute of Technology, under contract with the National 
Aeronautics and Space Administration.}, and other available
images at optical or X-ray wavelengths. 
In addition, we cross-correlated our data with publicly available
SZ catalogues from large surveys such as {\it Planck}, SPT,
and ACT. For a detailed description of the construction of the 
\rtwo\ galaxy cluster catalogue, we refer to B\"ohringer et al. 
(in prep.). For the compilation of the supercluster catalogue we used 
908 clusters discarding 11 clusters with missing redshifts.

\section{Construction of the REFLEX II supercluster catalogue}

In this section we describe the method to construct the supercluster 
catalogue from the \rtwo\ survey. This catalogue is the first supercluster 
catalogue based on a complete and homogeneous X-ray flux-limited cluster 
sample. The method to compile the supercluster catalogue is the 
friends-of-friends algorithm. This algorithm searches for neighbours of 
clusters within a certain radius, defined by the linking length $l$. It 
begins with the first cluster in the list of clusters and searches for other 
clusters (``friends'') in a sphere with the radius $l$ around each cluster. All 
clusters within this sphere are collected to the first system. A sphere 
around each friend is then searched for further clusters (``friends of 
friends'') in a recursive manner. All clusters connected in this way are 
assigned to a supercluster.

The crucial step in applying the friends-of-friends algorithm is the
specification of an appropriate linking length. If the linking length
is too small, only the cores of superclusters are selected. If it is
too large, individual clumps begin to grow together and build a larger 
system. As the number of clusters decreases significantly with redshift, 
one unique linking length for all clusters, which is normally used for 
volume-limited samples, would be either too large to detect nearby 
superstructures or too small to find superclusters at high redshifts.

The superclusters of the catalogue compiled by\defcitealias{zucca-93}{ZZSV}
\citet[hereafter ZZSV]{zucca-93}, were selected using a linking length 
depending on the \emph{overdensity} $f$, i.e. the cluster density enhancement 
over the mean cluster density, $f=n/n_0$. As the local density 
$n\varpropto l^{-3}$, the linking length $l\varpropto (fn_0)^{-1/3}$. 
The sample used by~\citetalias{zucca-93} was volume-limited, so the same 
linking length could be used for the whole sample independent of the cluster 
redshifts. As the density of the cluster sample varies with the position 
in the sky, the linking length was weighted with the inverse of the 
selection function that described the density variations. 

The comoving volume of a redshift shell is defined by 
\begin{equation}
  V = \frac{A}{3}\left[d_c(z_{\mathrm max})^3-d_c(z_{\mathrm min})^3\right]
\end{equation}
in units of Mpc$^3$ where $A=4.24$~sr is the the area of the
sky covered by the \reflex\ survey, and $d_c(z)$ is the comoving distance
at redshift $z$. The comoving mean cluster density in these redshift 
intervals is
\begin{equation}
  n_0=\frac{N}{V}
\end{equation}
in units of Mpc$^{-3}$, where $N$ is the number of clusters in the 
redshift interval. The average distance between two clusters in a given 
redshift shell is then $\bar{d}=n_0^{-1/3}$ and the linking
length is therefore
\begin{equation}
  l=(n_0f)^{-1/3}\,.
\end{equation}

\begin{figure}%[t]
  \centering
  \resizebox{\hsize}{!}{\includegraphics{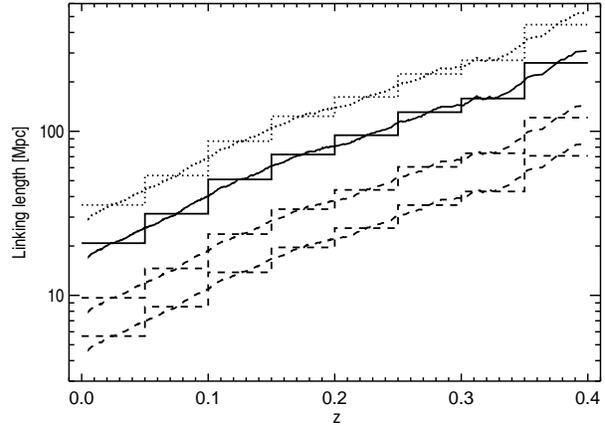}}
  \caption{
    Linking length as a function of cluster redshift.
    We compare the traditional linking length determined in discrete 
    redshift intervals (steps) to the continuous linking length determined
    at each cluster redshift over a range of the overdensity parameter, $f$. 
    A larger $f$ corresponds to a smaller linking length, i.e. 
    for the values of $f$ ranging from 2, 10 (solid line), 100, and 500
    the corresponding curve moves from top to bottom. 
  }
  \label{fig:llen}
\end{figure}

It is clear from Eqs. (1)--(3) that the linking length depends on the 
choices of the overdensity, $f$, and of $\Delta z$, i.e. the size of the 
sliding window in redshift, $\Delta z$=$z_{\mathrm {max}}$-$z_{\mathrm {min}}$. 
The larger the 
value of $f$, the higher the density peak we probe, which means that only 
the rare high density peaks are considered yielding a smaller linking 
length. Hence we expect to find a smaller number of superclusters. 
This effect is shown in 
Fig.~\ref{fig:llen}. The linking length with our reference value, $f$=10, 
is shown as a solid line, and $f=2$ corresponds to the dotted line above, and 
$f$=100 and 500 are shown as dashed lines. There are two curves corresponding 
to each overdensity parameter, which represent two different ways of calculating
the linking length. The step-like curves are produced by taking a $\Delta z$  
with a fixed z$_{\mathrm {max}}$ and z$_{\mathrm {min}}$, in this case 
$\Delta z$=0.05 in the 
redshift range covered by \rtwo\. The advantage of this choice is that 
within a redshift bin of 0.05, the linking length is fixed. Alternatively we 
devise a linking length that varies continuously with redshifts, which is 
constructed by calculating the linking length at the given cluster redshift 
where the volume for the density calculation is defined by $\pm \Delta z=0.025$
on each side of the cluster redshift. The resulting linking length is shown 
as continuous curves in Fig.~\ref{fig:llen}. 
While the use of fixed redshift bins results in an effectively lower
$f$ value at $z_{\mathrm min}$ in comparison to $z_{\mathrm max}$, the use of
a continuous linking length avoids this artificial bias. We therefore
prefer and adopt the latter method. By construction the continuous linking 
length goes roughly through the middle of each discrete step. 

The supercluster catalogue that we present is constructed with the continuous 
linking length with a couple of modifications. For the 
linking length between two clusters within a supercluster we take the mean 
of two linking lengths assigned to each cluster. This procedure produces 
exactly the same superclusters regardless of the choice of the first cluster 
selected to search for its friends. Similar to the concept of the Brightest 
Cluster Galaxy (BCG) we introduce the notion of the Brightest Supercluster 
Cluster (BSC), and our search starts from the brightest clusters in the 
\rtwo\ catalogue. Again this does not affect the properties of the final 
supercluster catalogue since the mean of two linking lengths is used for the 
construction.

The influence of $\Delta z$ on the resulting supercluster sample is shown in 
the top panel of Fig.~\ref{fig:noscl}. The reference redshift bin-width of 
$\Delta z$=0.025 is shown as solid line, and other values taken at 
0.01, 0.05, and 0.1 are shown as dotted lines. The very small differences in 
the number of 
superclusters recovered with different values of $\Delta z$ indicates that 
the linking length is not too sensitive to the choice of $\Delta z$ at a fixed 
overdensity parameter. 

Fig.~\ref{fig:noscl} also shows the number of superclusters obtained for 
different overdensities. Towards low overdensities individual superclusters at 
lower redshifts begin to grow together until they build mainly one large 
structure. Towards high overdensities the number of superclusters decreases 
due to the decreasing linking length. The maximum number of superclusters 
is at overdensities $f\approx4 \dots 10$. The main parameter which 
determines the nature of the superclusters is hence the overdensity parameter. 
In this paper we take $f$=10 as our reference for the argument given 
in the next section.

\subsection{Choice of the overdensity parameter}

To gain an understanding of the significance of the choice of a certain 
overdensity parameter, $f$, for the supercluster selection, we explore in 
this section the physical state of the superclusters as a function of $f$. 
For this consideration we assume a flat-$\Lambda$CDM universe with a matter 
density of, $\Omega_m = 0.27$. In this framework we can distinguish three 
characteristic evolutionary stages of bound structures: 
{\bf(i)} The structure is gravitationally bound, it has a mean density 
above about twice the critical density of the Universe, and the structure will 
gravitationally collapse in the future. {\bf(ii)} The structure has 
decelerated from the Hubble flow and is now starting to collapse.
This point is also called the moment of turn-around and occurs roughly 
at the time when the mean overdensity of the structure is about three 
times the critical density of the Universe. 
{\bf(iii)} The time of virialisation, which is the time when a homogeneous 
sphere approximation to the overdensity of the structure would collapse 
to a singularity. This stage is not reached by superclusters, since 
if they would have, they would become large galaxy clusters, by definition.

To classify superclusters according to this scheme, we thus need to know 
their mean matter overdensities with respect to the critical density of 
the Universe. This overdensity can not be determined easily and directly. 
Instead, we will use the cluster density as a proxy of the matter density. 
But in this application we have to keep in mind, that galaxy clusters show 
a biased density distribution, i.e. the density fluctuations in the 
cluster distribution are amplified by a bias factor with respect to the density
distribution of all the matter, which we define as 
$b_{\mathrm CL} = {\Delta_{\mathrm{CL},m} \over  \Delta_{\mathrm{DM},m}}$ where 
$\Delta_{\mathrm{X,m}} = {\rho_{X} -\rho_{\mathrm{m}} \over \rho_{\mathrm{m}}}$. 
The subscript $m$ refers to overdensities with respect to mean cosmic density.
The study of this bias for the case of the \rtwo\ sample shows that for 
the clusters at low redshift $z\leq0.15$ (which involves many clusters with 
low X-ray luminosity and low mass) bias factors between 3 and 4 should be 
expected, while for the higher luminosity at larger redshifts values of up 
to 5 should apply (Balaguera-Antolinez et al. 2011).

Thus for superclusters at low redshifts to be bound we find the following 
condition for case (i):

\begin{equation}
\Delta_{\mathrm{DM},c} + 1  = {(\Delta_{\mathrm{DM},m} + 1) \times \Omega_m} =  
\left( {\Delta_{\mathrm{CL},m} \over b_{\mathrm CL}} +1\right)  \times \Omega_m 
> 1~~,
\end{equation}

\begin{equation}
\Delta_{\mathrm{CL},m} > \left( \frac{1}{\Omega_m} -1 \right) b_{\mathrm CL} \sim 25 - 35~~,
\end{equation}
where the subscript $c$ denotes overdensities with respect to critical 
cosmic density, and for case (ii):
\begin{equation}
\Delta_{\mathrm{CL},m} > \sim 38 - 51,
\end{equation}
where $\Delta_{\mathrm{DM},c}$ and $\Delta_{\mathrm{DM},m}$ are the matter 
overdensities. $\Delta_{\mathrm{CL},m}$ is the cluster overdensity, and 
$b_{\mathrm CL}$ is the bias in the density fluctuations of clusters.

We therefore decided to explore the \rtwo\ superclusters with the above
results in mind for two bracketing values of $f$. With $f=10$ we sample
superstructures more comprehensively, slightly beyond definitely 
bound structures, and we expect that for these structures a minor part of 
matter may not collapse unlike the core part of the supercluster in the
future. We will explore this question of how much material of $f=10$ 
supercluster is actually gravitationally bound in more detail in a 
future paper by means of numerical simulations. With the other extreme
we study supercluster detected with $f=50$ which we expect to be
tightly bound and having started to collapse according to Eq. (6).

%Therefore, in the low redshift part of \rtwo\ we expect that a value of 
%$f \sim 10$ describes structures that are just marginally bound and a value of
%$f \sim 50$ describes structures that are tightly bound enough that they have 
%reached the moment of turn-around. 
Thus we adopted these two fiducial values
for the construction and discussion of our supercluster catalogue. 
In the higher redshift region the bias factor increases, which will 
result in lower matter overdensities for given value of $f$. But 
at the same time also $\Omega_m$ increases with redshift which partly 
compensates this effect.

\begin{figure}%[t]
  \centering
  \resizebox{\hsize}{!}{\includegraphics{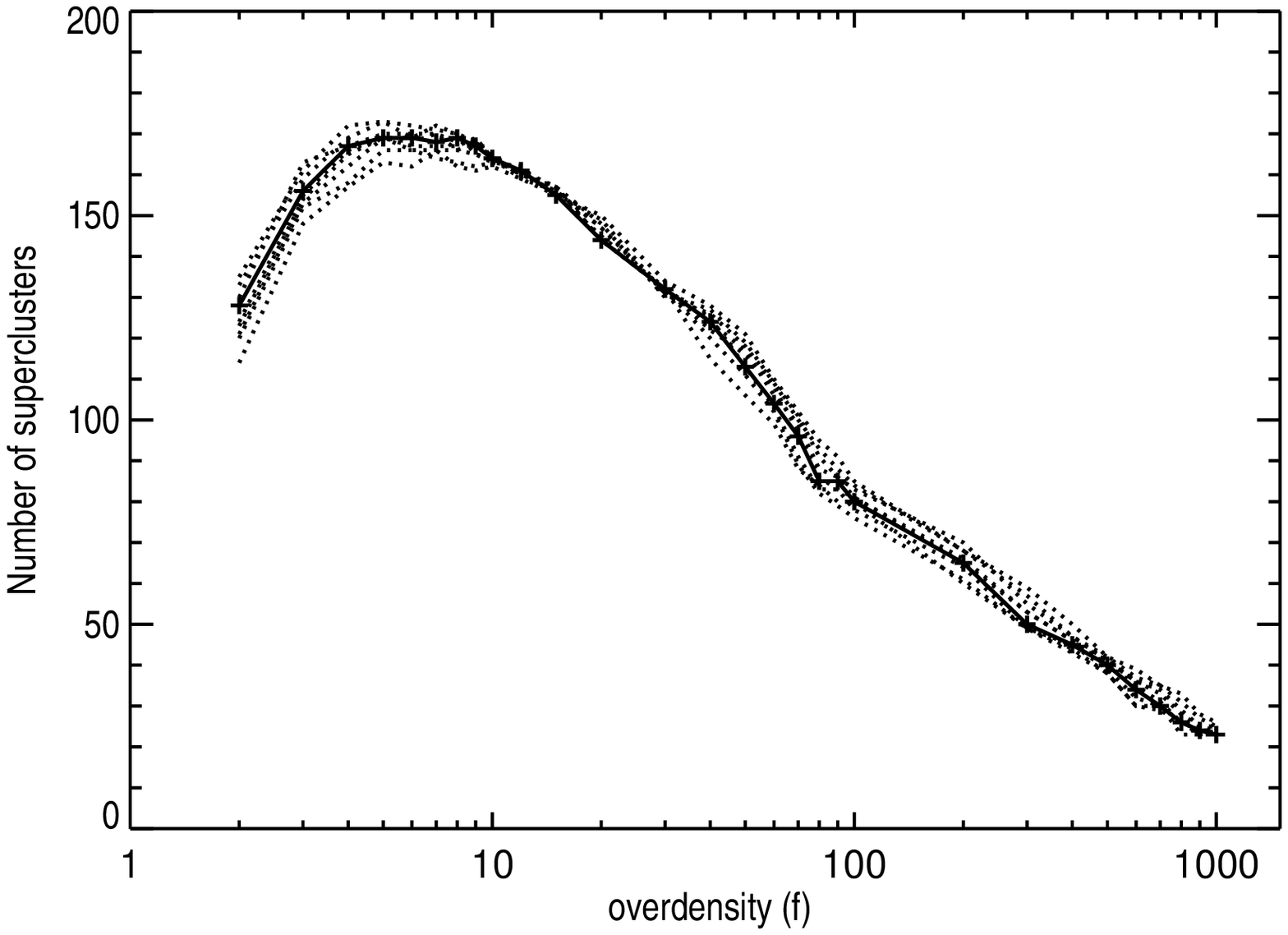}}
  \resizebox{\hsize}{!}{\includegraphics{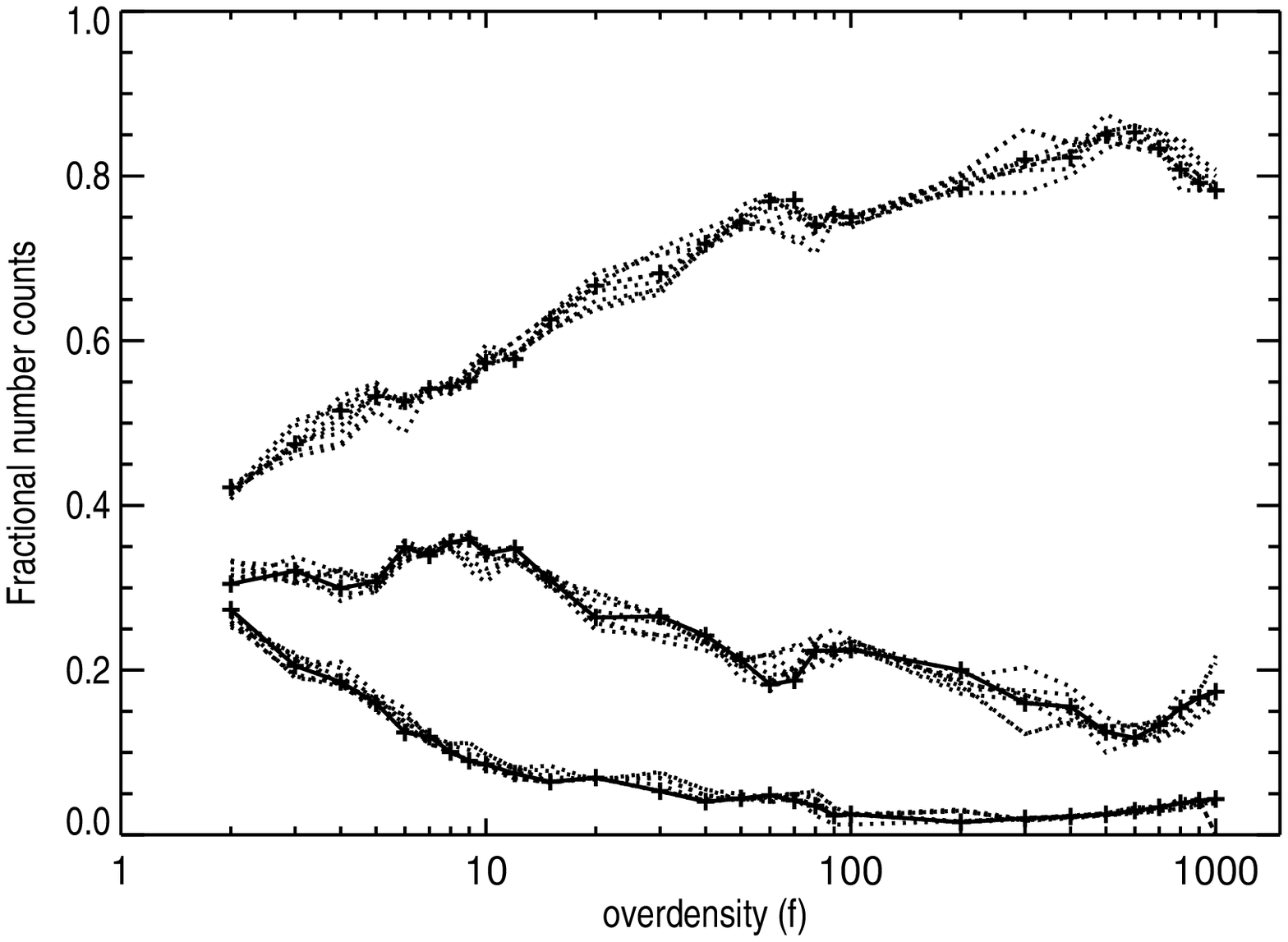}}
  \caption{({\bf Top}) Number of superclusters as a function of 
  the overdensity parameter, $f$. Our reference redshift window 
  $\Delta z$=0.025 is represented by a solid line with crosses, and other 
  dotted lines are for different redshift intervals, 0.01, 0.05 and 0.1. 
  ({\bf Bottom}) Fractional supercluster number counts as a function
  of $f$. Uppermost lines are the fractional counts for pair superclusters, 
  the group of lines in the middle is for the superclusters with between 3 to 
  5 cluster members and the lower most group of lines are for the rest of the 
  superclusters. The line scheme is identical to that of the upper panel.
  }
  \label{fig:noscl}
\end{figure}

The bottom panel of Fig.~\ref{fig:noscl} shows the fractional number of 
superclusters with different richness as a function of the overdensity 
parameter. The $f$=10 and $\Delta z$=0.025 case is shown with a solid line 
with crosses, and dotted lines represent different values of $\Delta z$,
0.01, 0.05 and 0.1. As in the top panel the choice of $\Delta z$ does not 
influence the number of superclusters as much as $f$. There are three groups 
of lines, the uppermost group is for the fraction of superclusters with two 
member clusters, the middle for the all superclusters with three member 
clusters, and the group at the bottom shows the rest of the richer 
superclusters.
At lower $f$ we find more rich superclusters since the linking 
length becomes very large, thus connecting physically un-connected clusters. 
As $f$ gets larger the fractional counts of pair superclusters, i.e.
superclusters with two cluster members, dominate more and more over other richer
superclusters. This is due to the fact that we are only probing the
highest peaks of the overdense regions with much smaller linking lengths. 
We note that the distribution of the fractional counts peaks at $f=10$ 
for the superclusters with three member clusters.
%Interestingly, we find that the fractional counts peak at $f=10$ for the
%superclusters with three member clusters.

\section{REFLEX II supercluster catalogue}

We present the \reflex\ supercluster catalogue constructed with $f=10$ in 
Table~\ref{tab:r2cat} in the Appendix. It comprises 164 superclusters 
constructed with 895 REFLEX II clusters at $z\leq0.4$ where 486 clusters are
found inside these superclusters. 
Besides coordinates and redshifts, we give the size of each supercluster,
$R_{\mathrm {max}}$, defined by half the maximum extent of a supercluster. 
We also indicate in the eighth column if superclusters constructed with 
$f=50$ are found in this catalogue. Given that the linking length 
for $f=10$ is larger than that for $f=50$, all superclusters built
with $f=50$ are found in $f=10$. They are either merged into a single 
supercluster or acquire more clusters at $f=10$.
The location in the sky and redshift of the superclusters is determined by 
the X-ray mass-weighted mean of the member clusters. 
We adopt the scaling relation used in~\cite{hans-scaling} where the 
X-ray mass, $M_X$, is proportional to $L_X^{0.62}$. 

\begin{figure}%[t]
  \centering
  \resizebox{\hsize}{!}{\includegraphics{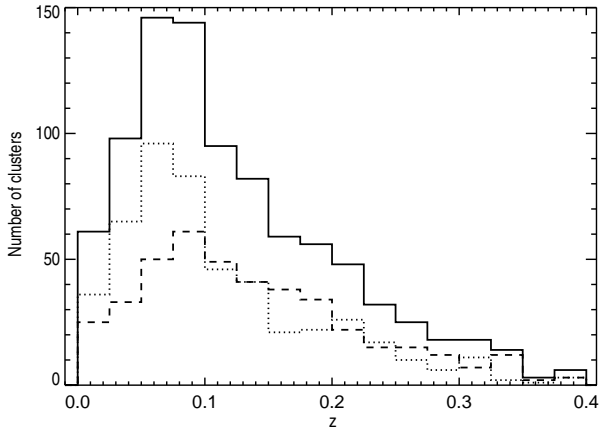}}
  \caption{
  Redshift distribution of \rtwo\ clusters (solid) compared to that
  of the clusters in superclusters (dotted line). The dashed line 
  represents the clusters that are not identified with superclusters in
  the \rtwo\ cluster catalogue.
  }
  \label{fig:count_z}
\end{figure}

At an overdensity of $f=10$ all known nearby superclusters such as Shapley, 
Hydra-Centaurus, Horologium-Reticulum, Aquarius and Pisces-Cetus are found. 
Some of them are split into sub-groups. This is due on the one hand to the 
small linking length at low redshifts and on the other hand to the fact that 
not all Abell clusters which are normally included in these superclusters,
like those listed for example in EETMA, are contained in the \rtwo\ 
sample, so that, even with larger linking lengths, not all subgroups
could be linked together. Choosing a just slightly lower or higher
overdensity like e.g. $f=8$ or $f=12$ does not make much difference.

Fig.~\ref{fig:count_z} compares the redshift distribution of \rtwo\ 
clusters shown in a solid line with that of members of superclusters
in a dotted and the field clusters in a dashed line. 
80\% of the \rtwo\ clusters lie below $z=0.2$ and 94\% below $z=0.3$.
The supercluster distribution practically follows the distribution of 
the \rtwo\ clusters, however, the number of clusters in superclusters
drops rapidly at $z\geq0.1$. In addition the number of clusters in the 
superclusters falls a bit faster than that of the field clusters. The 
distribution peaks slightly below $z$=0.1. The largest fraction of 
superclusters was found at this redshift together with another peak at $z$ around 
0.2. The cluster density 
decreases with redshift, hence fewer superclusters are found at high 
redshift. Due to the decrease of the cluster density at higher redshifts
$z\geq$0.3 and the consequently larger linking lengths 
the definition of the superclusters is less solid at higher redshifts. 
Therefore the interpretation of the supercluster properties at high redshifts, 
especially for those with two members with a relatively large distance
between them is rather speculative. Hence in this paper we restrict our 
analysis only up to
$z\leq0.4$. We further investigate the significance of these pair superclusters
later in this paper. In the next section we discuss the details 
of the supercluster properties.

\section{Properties of the superclusters}

\subsection{Specific superclusters}

\begin{figure}%[t]
  \centering
  \resizebox{\hsize}{!}{\includegraphics{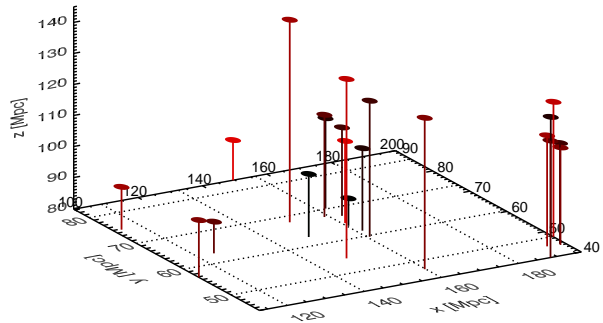}}
  \caption{
    Distribution of the individual \rtwo\ clusters in the Shapley 
    supercluster. 
    The \rtwo\ supercluster catalogue compiled with $f=10$ divides
    this supercluster into three sub-superclusters, and in total 
    there are 21 clusters associated.
    They are colour-coded such that the brighter red corresponds to 
    more luminous clusters, and darker red to the less luminous.
    The core of Shapley shown as the largest group in the middle 
    contains most number of brightest clusters.
  }
  \label{fig:shapley}
\end{figure}

In this section we identify some of the well-known superclusters 
in our catalogue. The spatial distribution of \rtwo\ clusters are
shown in Fig.~\ref{fig:scl-dist} in the Appendix. The clusters in the
superclusters are marked in blue dots linked by lines to their BSC, and 
the field clusters are marked in red dots.
The \emph{Shapley supercluster}, known as the richest supercluster at 
$z=0.046$, is found and according to the classical identification of the 
Shapley supercluster, it is divided into three sub-groups consisting of 21 
clusters in total in our supercluster catalogue compiled with $f=10$.
The three-dimensional positions of the individual clusters belonging to the
three sub-groups are shown in Fig.~\ref{fig:shapley}.
The clusters are coloured according to their luminosities, the
most luminous clusters are clustered in the main group in the middle. 
The left-most group on the x-axis is the supercluster ID 100 in Table~\ref{tab:r2cat}, 
the large group in the middle is ID 98, and six clusters at the lower 
right corner belong to ID 94. We note that all three regions grow together 
at $f=6$.

\emph{Hydra-Centaurus} is located in a similar region of the sky 
as the Shapley supercluster, but at a lower redshift, $z\sim0.011$.
In our catalogue it is divided into two superclusters, ID 92 and
95 with 12 clusters in total.

\emph{Aquarius B} has 16 cluster members in our sample, divided
in two superclusters, ID 148 and 157. This is remarkable
as it is at a redshift of $z\sim0.08$, and is nearly twice as far away 
as Shapley. This supercluster seems to be very rich in X-ray clusters,
but not extraordinarily rich in optical clusters (see EETMA).

ID 40 and 42 with 12 members altogether build the \emph{Horologium-Reticulum} 
supercluster at $z\sim0.07$ and $z\sim0.06$ respectively. This is the largest 
supercluster known in the southern hemisphere. It was first reported 
by~\citet{shapley-35} and studied by~\citet{lucey-83} 
and~\citet{chincarini-84}. This supercluster is very rich in Abell 
clusters (see for example EETMA) and still quite rich in X-ray clusters, 
although less rich compared to Shapley or Aquarius B due to its larger 
distance.

ID 120 has seven members and a redshift $z=0.0554$. It can be identified
with supercluster 172 in EETMA and supercluster 62/2 in
\cite{zucca-93}. It is also one of the richest superclusters 
in X-rays. ID 89 at $z=0.0848$ is with ten members also quite rich, which
is identified with supercluster 126 in EETMA.

The superclusters at redshifts $z>0.2$ are not as rich as the 
nearby superclusters, as the number density of clusters decreases 
rapidly towards higher redshifts while the volume gets larger. 
The number of superclusters with more than three clusters beyond $z>0.3$
is one, between $0.2<z<0.3$ is six, and 21 for $0.1<z<0.2$. The richest
superclusters above $z=0.1$ are ID 10 at $z=0.111$ with seven clusters and 
ID 85 at $z=0.128$ with six clusters. There are six superclusters with five 
or more member clusters above $z$=0.1.

\subsection{Multiplicity function}

The multiplicity function is the number of clusters found in a supercluster,
which is analogous to the richness parameter for an optical cluster.
We include in our catalogue already superclusters with only two members.
Of course there is the question if there is more to these cluster pairs
than just two clusters which are close together by chance. We attempt to
answer this question in section 6 where we investigate if cluster pairs
selected with a high X-ray luminosity threshold are found to be located
in superclusters which have a higher multiplicity when the X-ray luminosity cut
is lowered. Since the answer we find is encouraging, we take cluster pairs
into our catalogue expecting that the majority of these systems would be 
in richer systems if the selection was extended to lower mass cluster and 
galaxy group regime.

\begin{figure}%[t]
  \centering
  \resizebox{\hsize}{!}{\includegraphics{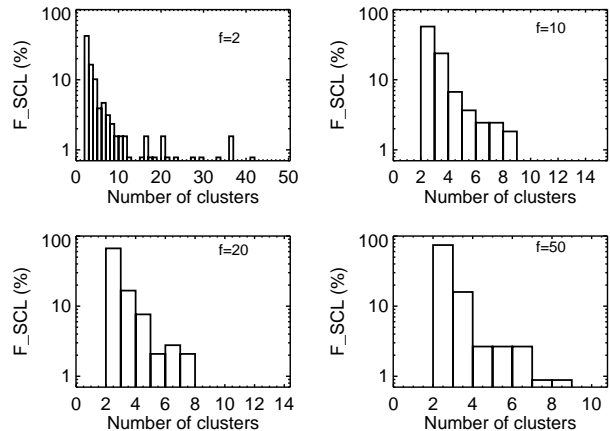}}
  \caption{
    Fractional multiplicity function for a range of overdensity
    parameters, $f$. The value of $f$ is drawn in each plot. In all cases
    pair superclusters dominate the distribution. As $f$ gets smaller,
    which corresponds to a large linking length, we find richer systems 
    (Upper left). The reverse is true that a smaller linking length only
    connects close-by systems, hence less rich systems are more common
    (Lower right). 
  }
  \label{fig:multi}
\end{figure}

The multiplicity of the \rtwo\ superclusters is given in 
Table~\ref{tab:r2cat}, and the fractional multiplicity functions are shown 
in Fig.~\ref{fig:multi} for a number of overdensity parameters. 
A large linking length corresponding to a small value of $f$ yields
a larger spread in the multiplicity distribution. This is expected 
since a large linking length collects, by construction, clusters 
lying above a smaller density threshold yielding larger and richer
superclusters. There are some extremely rich superclusters with more than
20 members in the top left panel of Fig.~\ref{fig:multi}.
For a smaller linking length corresponding to a larger overdensity, the 
distribution is tighter, steeply rising towards low multiplicities. 
This implies that this high overdensity threshold probes only the most 
dense areas, hence those superclusters are dominated 
by smaller pairs. At $f=50$ the largest multiplicity occurs
at the core of the Shapley supercluster. It still has a multiplicity of 6 at
$f=1000$.

The multiplicity function of our catalogue can also be understood from  
plot in the bottom panel of Fig.~\ref{fig:noscl}. 
With a small linking length the fraction of pair superclusters dominates
the total number of superclusters while the number of richer 
superclusters increases clearly with a larger linking length.

\begin{figure}%[t]
  \centering
  \resizebox{\hsize}{!}{\includegraphics{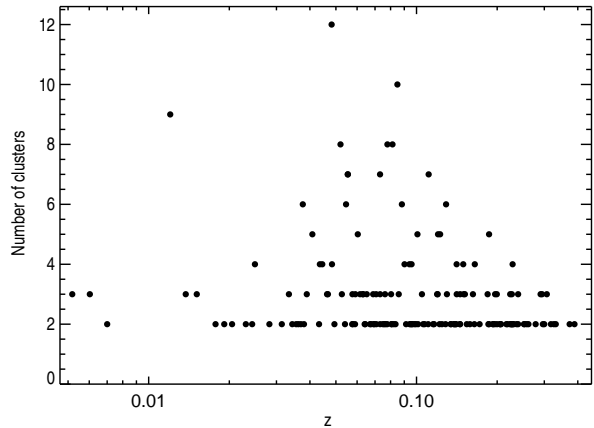}}
  \caption{
    Multiplicity as a function of redshift. 
    The pair superclusters are ubiquitous over the redshift range 
    while the richer systems tend to segregate around $z=0.1$.
  }
  \label{fig:nomem_z}
\end{figure}

Fig.~\ref{fig:nomem_z} shows the multiplicity distribution of superclusters 
as a function of redshift for $f=10$. The pair superclusters appear
at all redshifts while richer clusters tend to segregate slightly
below z=0.1. At $z\geq0.2$ nearly all superclusters are pair superclusters.
No superclusters with more than 5 members are found at redshifts $z>0.2$. 

\subsection{Extent of superclusters}

\begin{figure}%[t]
  \centering
  \resizebox{\hsize}{!}{\includegraphics{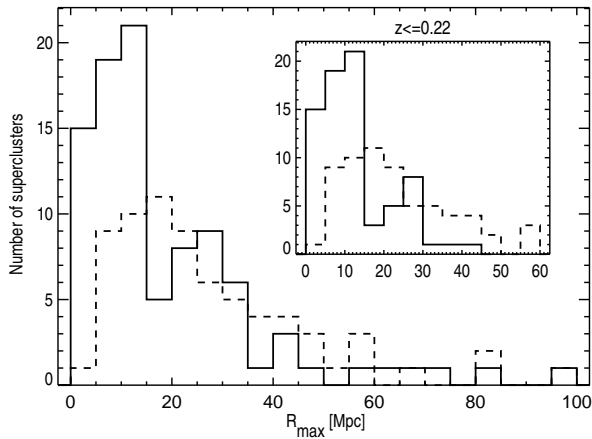}}
  \caption{
    Histogram of the maximum radius of the superclusters, $R_{\mathrm max}$.
    The superclusters with two cluster members are marked by a solid line, 
    and richer superclusters by a dashed line. The smallest radii 
    are predominantly found in pair superclusters. The inset plot
    shows the equivalent plot restricting the redshift range to
    $z\leq0.22$ where the largest radius is 56 Mpc. 
    See text in Sec. 6 for the choice of this specific redshift.
  }
  \label{fig:size}
\end{figure}

The distribution of the maximum radius, R$_{\mathrm {max}}$, defined as half the 
maximum separation between the member clusters, of the \rtwo\ superclusters 
is given in Fig.~\ref{fig:size}. 
The solid line denotes the distribution of the pair 
superclusters, and the dotted line is for the rest of richer superclusters. 
The distribution peaks around 10~Mpc, and the largest radius is just below 
100~Mpc for the entire supercluster sample. The smallest and the largest 
superclusters are dominated by pair superclusters while the medium-sized 
supercluster distribution is equally occupied by both pair and richer 
superclusters. Also shown in Fig.~\ref{fig:size} is an inset of the same 
distributions where the redshift is restricted to $z\leq0.22$. This particular
choice of the redshift cut will be explained later in Sec. 6, however, in the 
meantime we note that all eleven superclusters of size above 60~Mpc are located
at $z>0.22$.
We have started to investigate the corresponding supercluster properties in 
cosmological simulations, and here we only make a short remark that the 
distribution of the extent of \rtwo\ superclusters resembles that of the 
superclusters constructed from the simulation. We will defer a detailed 
discussion to a future publication (Chon et al., in prep.).

%With the maximum radius of the superclusters we can calculate in retrospect 
%what the overdensity of a supercluster is by defining a spherical volume 
%which includes the member clusters. It turns out that all superclusters but 
%one have an overdensity larger than 50.

\subsection{Comparisons with other catalogues}

A direct comparison with other supercluster catalogues is not possible as this 
is the first time that a complete flux-limited cluster catalogue was used for 
the supercluster selection. 
Table~\ref{tab:others} lists supercluster catalogues constructed directly 
with samples of galaxy clusters together with some of their properties. 
All catalogues except EETMA are 
exclusively based on Abell or Abell/ACO clusters. The early supercluster 
catalogues are based on relatively small cluster samples, as only a small 
fraction of the Abell clusters had a measured redshift. In many cases clusters 
with estimated redshifts were also included in the samples. \citet{postman-92} used a sample of Abell (1958) and ACO (1989) clusters with only
15 clusters from the ACO catalogue resulting in a less complete sampling
of the southern sky. ~\citet{cappi-92} performed a similar analysis 
with a more equal coverage of the northern and southern sky, confirming the 
results of~\cite{postman-92} in the north.

The superclusters of all catalogues except those of Abell's (Abell 1961)
were selected using some kind of friends-of-friends analysis. 
Abell found superclusters 
by cluster counts in cells with a projected galaxy distribution for each 
distance class. In five catalogues the linking length depends on the 
overdensity and these catalogues were compiled for different overdensities, 
whereas for the other catalogues fixed linking lengths were used.

EETMA were the first to use X-ray selected cluster samples. Their goal was to 
study the clustering properties of X-ray clusters in comparison to Abell 
clusters. Four different X-ray cluster samples were used: the all-sky \rosat\ 
bright survey \citep{voges-99}, a flux-limited sample of bright clusters from 
the southern sky~\citep{degrandi-99}, the \rosat\ brightest cluster sample 
from the northern sky~\citep{ebeling-98} and a number of \rosat\ pointed 
observations of the richest ACO clusters~\citep{david-99}. 
The mixture of different samples produces an effective sample that is neither 
complete nor homogeneous. For the comparison of the clustering properties with 
Abell clusters this might be nevertheless a good approach. Furthermore using 
different samples simultaneously was at that time the only possibility to get 
a relatively large number of clusters, which is necessary in order to compare 
it with the large number of Abell clusters. EETMA give a list of 99 Abell 
superclusters that contain X-ray clusters of galaxies, thereof 47 
superclusters in the southern hemisphere. 34 \rtwo\ superclusters could be 
clearly identified with superclusters in that sample. Among ten additional 
superclusters in the non-Abell X-ray clusters by EETMA three could be 
identified with the \rtwo\ superclusters. A number of superclusters in 
\rtwo\, especially the very rich ones, can also be identified with 
superclusters in other catalogues that cover the southern sky, 
e.g.~\citetalias{zucca-93} or~\citet{kalinkov-95}.

\begin{table*}
\centering
\caption{A list of supercluster catalogues compiled in the past using
  Abell/ACO and X-ray cluster catalogues. For each catalogue the number
  of clusters, the catalogue on which the supercluster search is based
  the number of clusters with a measured or estimated redshift, the
  linking length $l$ in units of $h^{-1}$~Mpc except this work, the 
  overdensity $f$ and
  the number of identified superclusters are listed. Additionally the
  selection criteria (distance group $D$, richness $R$, redshift $z$,
  distance $d$, magnitude of tenth brightest galaxy $m_{10}$ or
  galactic latitude $b_{II}$) are given.  Either the linking length $l$
  or the smallest overdensity $f$ used to identify the superclusters
  are specified. Catalogues including superclusters in the southern
  hemisphere and all-sky catalogues of superclusters were only compiled
  after the ACO catalogue had been published. \citetalias{zucca-93}
  used only Abell/ACO at distances $d\lesssim300$~Mpc/h. The supercluster 
  catalogue compiled by EETMA using Abell/ACO clusters is an update of the
  catalogue of~\citet{einasto-97}, and this is again an update of
  \citet{einasto-94}. EETMA used four different catalogues of X-ray
  clusters with different flux limits. 99 of their Abell/ACO
  superclusters contain at least one X-ray cluster, additionally 19
  superclusters containing exclusively X-ray clusters were found.}
\begin{tabular}{lrlcrrrl}
\hline\hline
 & & & \multicolumn{1}{c}{redshift} & & & & \\ 
Author & \#Cl & Catalogue & meas./est. & \multicolumn{1}{c}{$l$} & $f$ & \multicolumn{1}{c}{\#SC} & Cluster selection criteria\\
\hline
\citet{abell-61} & 1682 & Abell & 0/1682 & & & 17 & \\ \hline
\citet{rood-76} & 27 & Abell & 27/0 & 25 & & 5 & $D=0\dots2$\\
                & 83 & Abell & 28/0 & 25 & & 15 & $D=3$\\
                & 166 & Abell & 8/0 & 25 & & 24 & $D=4$\\ \hline
\citet{thuan-80} & 77 & Abell & 77/0 & 36 & & & $D\leq4$, $z\leq0.08$\\ \hline
\citet*{bahcall-84} & 104 & Abell & 104/0 & & 20 & 16 & $D\leq4$, $R\geq1$, $z\leq0.1$\\ \hline
\citet*{batuski-85}  & 652 & Abell & 307/345 & 40 & & 102 & $R\geq0$, $z\leq0.13$\\ \hline
\citet{west-89} & 286 & Abell & 286/0 & 25 & & 48 & $R\geq0$, $z\leq0.1$\\ \hline
\citet{postman-92} & 351 & Abell/ACO & 351/0 & & 2 & 23 & $R\geq0$, $m_{10}\leq16.5$\\ \hline
\citet*{cappi-92} & 233 & Abell/ACO & 219/14 & & 1.9 & 24 & $R\geq0$, $m_{10}\leq16.5$, $|b_{II}|\geq40^\circ$ \\ \hline
\citetalias{zucca-93} & 4319 & Abell/ACO & 1074/3245 & & 2 & 69 & $d\lesssim300$h$^{-1}$~Mpc, $|b_{II}|>15^\circ$\\ \hline
\citet*{kalinkov-95} & 4073 & Abell/ACO & 1200/2873 &  & 10 & 893 & $z\leq0.2$\\ \hline
\citet{einasto-94,einasto-97} & 1304 & Abell/ACO & ~870/~434 & 24 & & 220 & $R\geq0$, $z\leq0.12$\\ \hline
\citetalias{einasto-01} & 1663 & Abell/ACO & 1071/592 & 24 & & 285 &$R\geq0$, $z\leq0.13$\\
                        & 497 & X-ray & 497/0 & 24 & & 19 & $z\leq0.13$\\ \hline
%\citet{einasto-07} % this is galaxy based, not cluster based &  \\ \hline
this work & 908 & \rtwo\ & 908/0 & & 10 & 164 & $f_\mathrm{X}\geq1.8\times10^{-12}$~erg s$^{-1}$ cm$^{-2}$ \\ 
\hline
\end{tabular}
\label{tab:others}
\end{table*}

The fact that a significant number of superclusters are common in
these catalogues constructed with significantly different selection
functions shows that many of these superclusters are very characteristic
and prominent structures that clearly stick out of the general cluster
distribution. This is analogous to the above finding that very similar
superclusters are found for a wide range of linking length.

\section{Volume-limited sample}

To overcome the problem of a supercluster catalogue in which the linking 
length varies with redshift, and in which consequently the nature of the 
superclusters also changes with distance, we also considered the compilation 
of a supercluster catalogue based on a volume-limited sample (VLS) of \rtwo\  
clusters. This sample has been constructed by a means of constant luminosity 
and redshift limit. Thus volume-limited samples of \reflex\ could be any group 
of clusters with a luminosity and redshift limit not crossing the parabolic 
boundary in the $L_\mathrm{max}$--$z$ plane. Therefore the dynamical state 
of the superclusters from the VLS of clusters does not change with redshift. 
Its limitation lies in, however, small number statistics.

The VLS also gives us the chance to explore the robustness of the selection 
of superclusters for various selection criteria. 
In relation to the construction of the supercluster catalogue in the previous 
sections our most urgent question is the following. When going to higher 
redshifts, which corresponds to higher luminosity limits, the superclusters are 
selected from a density distribution with decreasing mean density and a larger 
linking length. 
Thus it would be important to know if we would have selected 
the same superclusters had we more data with a lower luminosity limit. 
With the VLS we can address this question by ``artificially'' 
increasing the luminosity limit (reducing the information) and by comparing 
the supercluster samples at increasing luminosity cuts. 

The other question arises from the definition of superclusters. It is 
somewhat arbitrary in the sense that superclusters are not collapsed objects
and their extent cannot be defined uniquely. This means that understanding 
the contamination in the catalogue is important. To answer this question we 
need a simulation to which we can apply the same selection criteria as to our 
\reflex\ sample. In this paper, however, we take an alternative approach that 
utilises the \reflex\ data already in hand, and defer further discussions 
based on simulations to the future publication (Chon et al., in prep.). 

By constructing hierarchical layers of VLS 
with increasing lower luminosity cuts we will address the questions 
raised above. As a reference VLS we cut out a volume defined
by $z\leq0.1$ and $L_{\mathrm X,0}\geq5\times10^{43}$ erg/s. This is one of the 
largest volume-limited samples that can be constructed from \rtwo\. 
With this sample
we achieve a relatively high cluster density rather than the largest volume.
There are in total 171 \rtwo\ clusters, and we find 31 superclusters
with $f$=10. Their properties are given in Table~\ref{tab:vls}. 
The column headings are identical to Table~\ref{tab:r2cat} in the Appendix
except that we identify the newly found superclusters with the superclusters
in the full \rtwo\ supercluster sample in Table~\ref{tab:r2cat} 
in the eighth column.

\begin{table*}
\centering
\caption{List of the volume-limited superclusters. 
  This is the catalogue, namely $L_{\mathrm X,0}$ defined in Table~\ref{tab:pars}. 
  The columns are (1) Supercluster name
  (2) R.A. (deg) (3) Dec. (deg) (4) redshift (5) $R_{\mathrm max}$ (Mpc) 
  (6) Total luminosity ($10^{44}$ erg/s) (7) Multiplicity (8) Corresponding supercluster ID
  in Table B.1 (9) $f=50$ %(10) ID number. 
  The position and redshifts are mass-weighted by the member 
  clusters as described in the text. 
  RA and Dec are for J2000. $R_{\mathrm max}$
  is half the maximum extent between clusters in the supercluster.
  $L_{\mathrm X}$ is the sum of all member cluster luminosities. 
  The multiplicity is the number of member clusters. (8) lists the ID of
  the full \rtwo\ superclusters. $+1$ means that the superclusters
  in the volume-limited sample acquired a new cluster member.
  (9) Y marks the existence of the supercluster in the $f=50$ catalogue. 
  $Y-x$ indicates that $x$ number of clusters that 
  were missing in the $f=50$ catalogue and $Y/x$ indicates that
  the supercluster was divided into $x$ number of superclusters.
}
\label{tab:vls}
\begin{tabular}{c c c c c c c c c}
\hline
\hline
\multicolumn{1}{c}{Supercluster}& 
\multicolumn{1}{c}{$\alpha$[$^\circ$] } & 
\multicolumn{1}{c}{$\delta$[$^\circ$] } & 
\multicolumn{1}{c}{$z_\mathrm{SC}$} & 
\multicolumn{1}{c}{$R_\mathrm {max}$ [Mpc]} & 
\multicolumn{1}{c}{$L_X$ [$10^{44}$ erg/s]} & 
\multicolumn{1}{c}{Multiplicity} & 
\multicolumn{1}{c}{Table B.1 ID} & 
\multicolumn{1}{c}{$f=50$} \\ 
%\multicolumn{1}{c}{ID} \\
(1) & (2) & (3) & (4) & (5) & (6) & (7) & (8) & (9) \\ %& (10)\\
\hline
\hline
RXSCJ2357-3000 & 359.459 &  -30.005 &  0.0629 & 19.32 &  1.643 &  2 &4+1   \\%&         1 \\
RXSCJ0013-1915 &   3.454 &  -19.262 &  0.0943 &  2.66 &  1.922 &  2 &7     &Y\\%&        2 \\
RXSCJ0006-3414 &   1.669 &  -34.235 &  0.0480 & 11.48 &  2.170 &  2 &2     &Y\\%&        3 \\
RXSCJ0055-1341 &  13.851 &  -13.689 &  0.0550 & 28.15 &  7.992 &  4 &13,18+1&Y-2\\%&  4 \\
RXSCJ0321-4345 &  50.309 &  -43.765 &  0.0707 & 34.23 &  6.096 &  4 &39,40 &Y-2\\%&   5 \\
RXSCJ0357-5702 &  59.325 &  -57.049 &  0.0593 & 27.98 &  8.504 &  4 &6,42 &Y-1\\%&    6 \\
RXSCJ0435-3830 &  68.809 &  -38.509 &  0.0516 & 19.03 &  1.249 &  2 &	  \\%&	   7 \\
RXSCJ0429-1104 &  67.495 &  -11.070 &  0.0356 & 16.41 &  3.015 &  2 &50+1& \\%&       8 \\
RXSCJ0444-2109 &  71.083 &  -21.163 &  0.0698 & 12.48 &  1.701 &  2 & 52& \\%  &      9 \\
RXSCJ0549-2111 &  87.402 &  -21.192 &  0.0958 & 21.88 &  4.739 &  4 & 61&Y/2\\%&	  10 \\
RXSCJ0618-5519 &  94.621 &  -55.325 &  0.0539 & 24.65 &  4.013 &  4 & 62+1&Y-1\\%&   11 \\
RXSCJ0913-1044 & 138.279 &  -10.749 &  0.0541 &  6.57 &  6.743 &  2 & 68&Y\\%&   	  12 \\
RXSCJ1027-0945 & 156.961 &   -9.759 &  0.0614 & 18.83 &  1.569 &  2 & 77+1&\\%&      13 \\
RXSCJ1144-1504 & 176.072 &  -15.079 &  0.0722 & 12.85 &  1.230 &  2 & 84&\\%&        14 \\
RXSCJ1152-3259 & 178.138 &  -32.987 &  0.0691 & 11.28 &  1.497 &  2 & 86&Y\\%&   	  15 \\
RXSCJ1236-3435 & 189.189 &  -34.591 &  0.0770 & 13.21 &  1.326 &  2 & 90&\\%&        16 \\
RXSCJ1318-3055 & 199.529 &  -30.923 &  0.0487 & 39.00 & 16.593 & 11 & 94,98,100&Y/2-1\\%&   17 \\
RXSCJ1310-0218 & 197.746 &   -2.309 &  0.0845 & 49.43 & 13.863 &  8 & 89+1&Y/2-2\\%& 18 \\
RXSCJ1614-8311 & 243.631 &  -83.194 &  0.0731 &  8.24 &  3.248 &  2 & 112&Y\\%&      19 \\	 
RXSCJ1704-0115 & 256.061 &   -1.263 &  0.0914 &  3.98 &  2.713 &  2 & 116&Y\\%&      20 \\	 
RXSCJ2009-5513 & 302.419 &  -55.231 &  0.0549 & 33.32 &  7.533 &  5 & 120+1&Y-3\\%&  21 \\ 
RXSCJ2036-3505 & 309.066 &  -35.088 &  0.0908 & 10.52 &  5.741 &  3 & 134&Y\\%&      22 \\	 
RXSCJ2226-6452 & 336.711 &  -64.873 &  0.0937 & 29.83 &  5.989 &  4 & 23,156&Y-1\\%& 23 \\
RXSCJ2152-5641 & 328.064 &  -56.693 &  0.0780 & 30.95 &  7.511 &  6 & 145&Y-3\\%&    24 \\   
RXSCJ2211-0727 & 332.787 &   -7.464 &  0.0854 & 46.67 & 13.247 & 10 & 148,157&Y/2\\%& 25 \\
RXSCJ2147-4449 & 326.843 &  -44.818 &  0.0608 &  7.33 &  1.411 &  2 & 144&Y\\%&   26 \\
RXSCJ2152-1937 & 328.083 &  -19.625 &  0.0951 &  4.13 &  2.762 &  2 & 150&Y\\%&   27 \\
RXSCJ2306-1319 & 346.726 &  -13.325 &  0.0670 &  4.79 &  1.150 &  2 & 160&Y\\%&   28 \\
RXSCJ2316-2137 & 349.030 &  -21.623 &  0.0858 & 16.87 &  3.517 &  3 & 162&Y\\%&   29 \\
RXSCJ2318-7350 & 349.648 &  -73.850 &  0.0982 &  5.29 &  1.607 &  2 & 163&Y\\%&   30 \\
RXSCJ2348-0702 & 357.161 &   -7.048 &  0.0777 & 19.08 &  3.298 &  2 & &\\%&   31 \\
\hline
\end{tabular}
\end{table*}

In addition to the reference sample we construct another three volume-limited 
samples with increasing lower luminosity cuts, 1, 2, and 3$\times10^{44}$ erg/s.
Table.~\ref{tab:pars} summarises for the different volume-limited samples 
our findings. As expected the number of superclusters decreases as the 
lower luminosity limit increases. 

\begin{table*}
\centering
\caption{Parameters for the four volume-limited samples.
They are constrained by $z\leq0.1$ with the luminosity cut
given by L$_{\mathrm X,min}$ in the second column. The total
number of \rtwo\ clusters in each volume is shown
in the third column together with the number of superclusters
in the fourth. The last column lists the linking length used
to build the supercluster catalogue. 
}
\begin{tabular}{c c c c c}
\hline
\hline
\multicolumn{1}{c}{Sample}& 
\multicolumn{1}{c}{L$_{\mathrm X,min}$ [10$^{44}$ erg/s]} & 
\multicolumn{1}{c}{No. \rtwo\ clusters} & 
\multicolumn{1}{c}{No. Superclusters} & 
\multicolumn{1}{c}{Linking length [Mpc]} \\
%\multicolumn{1}{c}{Corresponding redshift}\\
(1) & (2) & (3) & (4) & (5) \\
\hline
\hline
L$_{\mathrm X,0}$ & 0.5 & 171 & 31 & 39.3 \\%& 0.093\\
L$_{\mathrm X,1}$ & 1.0 & 75 & 17 & 51.7 \\%& 0.128 \\
L$_{\mathrm X,2}$ & 2.0 & 24 & 6 & 75.6 \\%& 0.183 \\
L$_{\mathrm X,3}$ & 3.0 & 13 & 2 & 92.7 \\% & 0.22 \\
\hline
\end{tabular}
\label{tab:pars}
\end{table*}

\begin{figure}%[t]
  \centering
  \resizebox{\hsize}{!}{\includegraphics{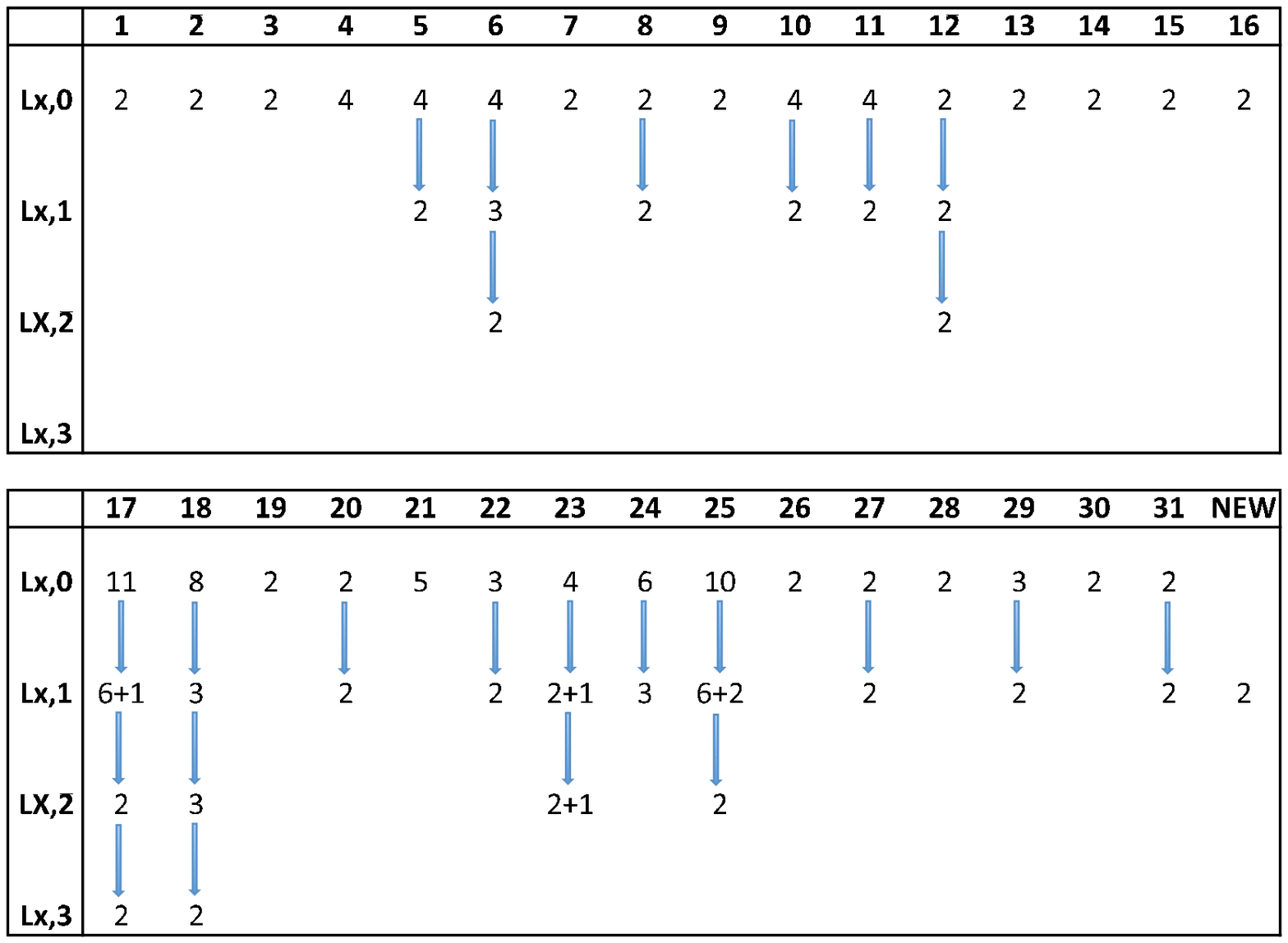}}
  \caption{
    Overview of the superclusters found with the four volume-limited samples
    of the \rtwo\ clusters. The four samples are constructed with the
    common redshift limit of $z\leq0.1$ and $L_{\mathrm {X,0,1,2,3}}$=0.5, 1, 2, and
    3 $\times 10^{44}$ erg/s. The numbers running from 1 to 31 indicate
    a unique ID for a supercluster, which corresponds to the superclusters
    in Table. 2. At the lowest luminosity cut 31 superclusters are identified,
    and as this cut becomes larger fewer superclusters survive.
    Blue arrows indicate that the supercluster found in the preceding
    lower luminosity cut survives to the next higher luminosity cut. 
    The numbers in the main part of two tables indicate the multiplicity of
    superclusters. The multiplicity is expected to be lower towards a higher
    luminosity cut as it is found for all cases. '+' sign indicates that the 
    supercluster
    obtained new member clusters at the given luminosity cut compared 
    to the preceding cut. One exception is shown in the last column of 
    the lower table where a new supercluster appears at the $L_{X,1}$ cut.
  }
  \label{fig:tree}
\end{figure}

To understand how each supercluster survives a larger luminosity cut
at each step we visualise this process by a tree structure shown in
Fig.~\ref{fig:tree}. There are two sub-tables, in which the first row
records the supercluster ID as in Table~\ref{tab:vls}. The rest
of the rows then contain the number of member clusters in each supercluster
if the supercluster was found in the specific volume-limited sample
indicated by the first column. The blue arrow denotes that the supercluster
from the lower luminosity cut survives up to the next higher luminosity cut.
A ``$+$'' denotes that the supercluster member gained an extra $x$ number of 
clusters that were previously not found within the same supercluster in 
the preceding lower luminosity cut. In summary there are 17, 6, and 2 
superclusters found in the $L_{\mathrm X,1}$, $L_{\mathrm X,2}$, and $L_{\mathrm X,3}$ 
samples respectively. As expected all superclusters identified at the higher
luminosity cuts are found in the catalogue compiled with the lower luminosity
cut with one exception, which is denoted as NEW in the last column of
Fig.~\ref{fig:tree}. This supercluster contains two member clusters at
a maximum radius of 23 Mpc. Hence we take this as an indicator for a possible 
contamination in our sample. 

What is interesting is then how this lower 
luminosity limit translates into the linking length as a function of redshift 
for the full sample. For each sample we have a fixed linking length, listed 
in column 5 of Table.~\ref{tab:pars}. By combining this table and 
Fig.~\ref{fig:llen} we can infer that the largest lower luminosity cut of the 
four volume-limited samples corresponds to a redshift of 0.22. This means that 
the statistics that we learn from the volume-limited samples can be applied
to the discussion of the full sample up to $z$=0.22. 
Therefore we interpret the emergence of a new, possibly unreal supercluster
in the $L_{\mathrm X,1}$ sample of 17 superclusters as a maximum contamination 
of 5.8\% for those 86\% of the superclusters up to $z$=0.22. 
We can also test how many superclusters found at redshifts $z\leq0.22$ 
are larger than 93 Mpc in diameter, and how many cluster members
they contain. We find four superclusters in this range, and they are 
richer systems with 3 or more member clusters. Hence our sample below
$z$=0.22 does not suffer from this random spatial coincidence by not
having a large pair supercluster. 
There is no good way to extrapolate this number out to $z$=0.4, however, 
we do not expect a much larger contamination in the rest 14 \% of the
catalogue.

\subsection{Volume fraction of superclusters}

The \rtwo\ supercluster catalogue includes 53\% of the \rtwo\ clusters, and 
in the VLS 62\% of the clusters are associated to superclusters. We have a hint 
from the spatial distribution of superclusters in Fig.~\ref{fig:scl-dist} 
that it is inhomogeneous. To test how much volume the superclusters 
take up, we calculate the volume that the superclusters occupy compared to 
the rest of the volume. Defining the volume of a supercluster is not unique 
especially for those with only pair clusters. Here we assume that superclusters 
occupy a spherical volume defined by the maximum radius of a supercluster as 
given in Tables~\ref{tab:r2cat} and~\ref{tab:vls}. The ratio of the volumes 
between the superclusters and the field is of 2\% for the reference VLS, 
and 0.6\% for the full supercluster catalogue. This is another indication 
that the environment occupied by the superclusters is significantly 
different from the field.

\section{X-ray luminosity distributions}

A census of the galaxy cluster population is provided by the cluster
mass function. This function can be predicted on the basis of cosmic
structure formation models such as the hierarchical clustering model for
CDM universes by Press-Schechter type theories~\citep{bond-91,press-74}. 
Due to the close relation of X-ray luminosity and gravitational mass of 
clusters the mass function is reflected by the X-ray luminosity function 
(XLF) of clusters. 

Thus we can use our data to study the observed XLF as a substitute for the
mass function. As already mentioned in the introduction, we expect the growth
of structure to continue in the dense supercluster region at a faster rate
than in the field in the recent past. Thus the mass function should be more
evolved. This evolution is most importantly characterised by a shift of the high
mass cut-off of the mass function to higher cluster masses. We therefore 
expect a more top-heavy mass function and XLF for the clusters in superclusters.

The XLF of the \rtwo\ clusters was derived by~\cite{reflex2} using the 
formula for the binned XLF 

\begin{equation}
\frac{dn(L_X)}{dL_X}=\frac{1}{\Delta L}\sum_{i=1}^N\frac{1}{V_{\mathrm{max}}(L_i)}\,,
\end{equation}
where $\Delta L$ is the size of the bin and
$V_\mathrm{max}=A/3 r^3_\mathrm{max}$ is the survey
volume, i.e. the volume defined by the survey area $A=4.24$~sr
and the maximum luminosity distance $d_L^\mathrm{max}=r_{max}(1+z)$,
at which a cluster with a given luminosity can just be detected. 
Note that we have a model for $r_{\mathrm max}$ which depends on the sky position
to account for the varying depth of the \rosat\ survey. 
The relationship between the survey volume and the X-ray luminosity 
$L_\mathrm{X}=4\pi d_L^2k(z)f_\mathrm{X}$ is
\begin{equation}
  V_\mathrm{max}(z)=\frac{A}{3}\left(\frac{1}{1+z}
\sqrt{\frac{L_\mathrm{X}}{4\pi k(z)f_\mathrm{X, lim}}}\right)^3\,,
\end{equation}
where $k(z)$ is the $k$-correction.

\begin{figure}%[t]
  \centering
  \resizebox{\hsize}{!}{\includegraphics{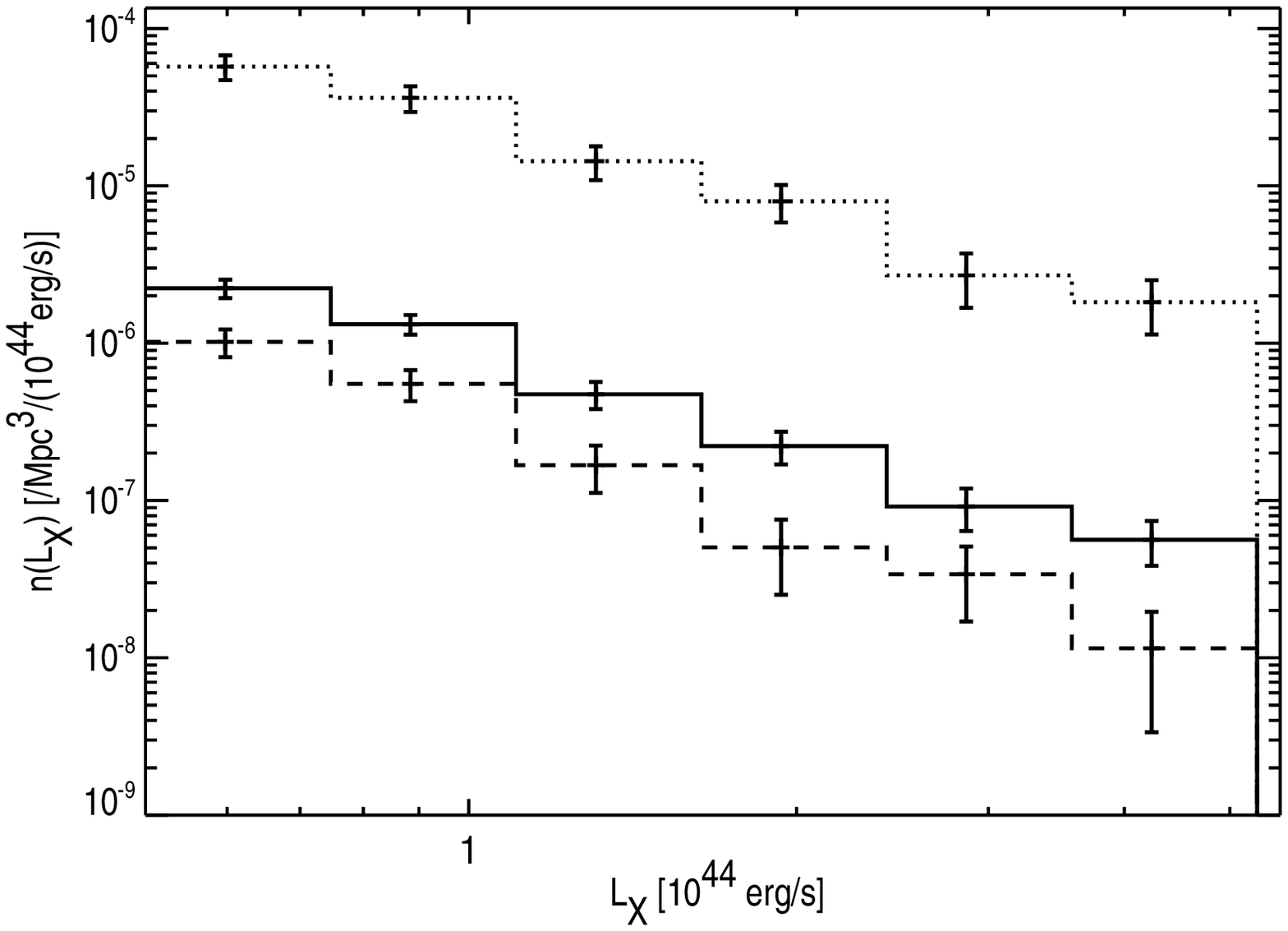}}
  \resizebox{\hsize}{!}{\includegraphics{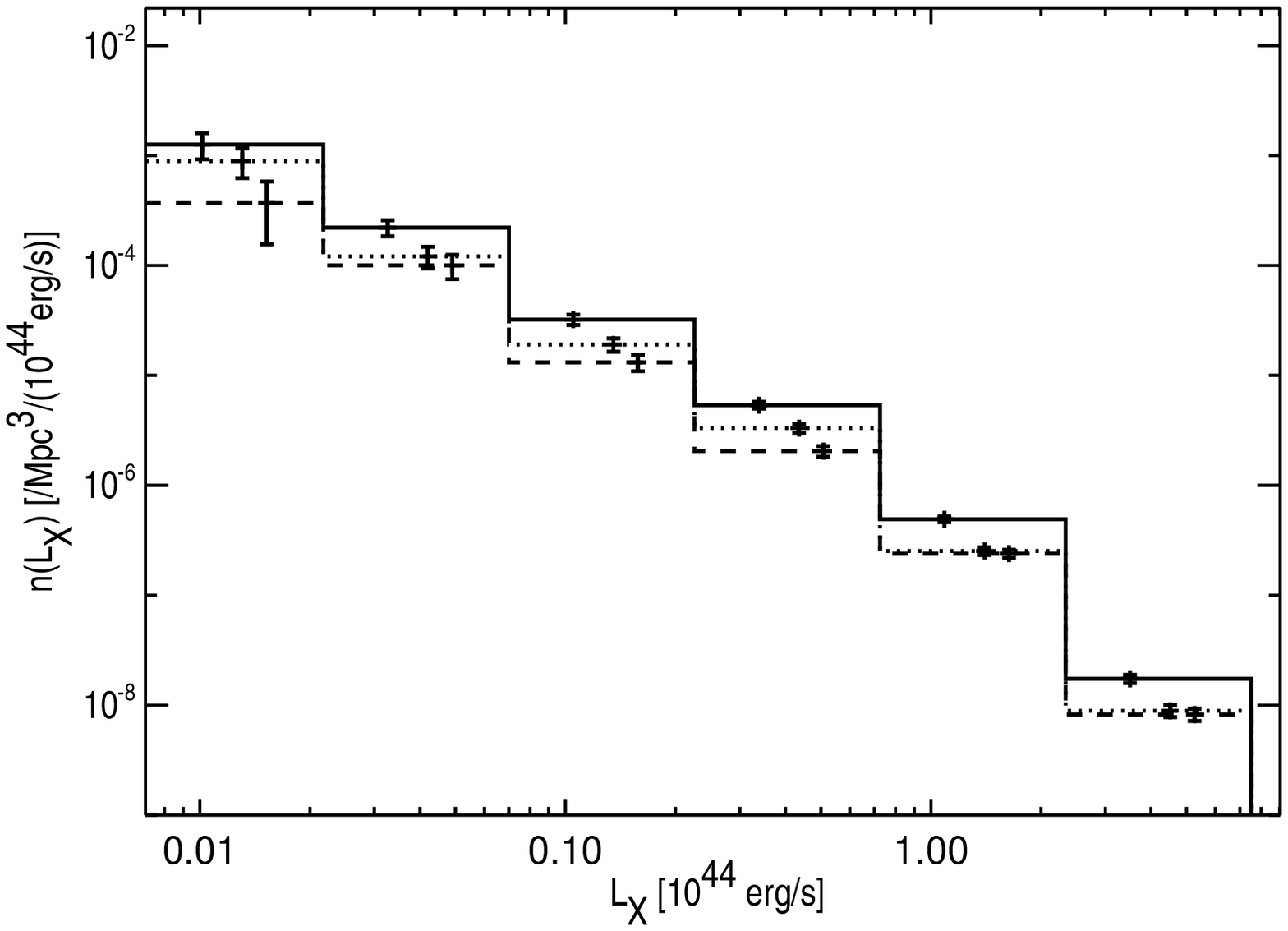}}  
  \caption{
    {\bf (Top)} 
    Differential luminosity function of the reference volume-limited 
    \rtwo\ clusters. The volume is defined by $L_{\mathrm X}>$ 
    5$\times 10^{43}$erg/s, and $z\leq0.1$. 
    The luminosity function of the clusters in the \rtwo\ catalogue 
    is denoted by the solid black line, that of the supercluster
    member clusters in a dotted line, and the field clusters are 
    marked by a dashed line.
    {\bf (Bottom)} Differential luminosity function of the flux-limited 
    \rtwo\ cluster sample.
    The line scheme is same as in the upper panel, and the errors 
    are displaced slightly for clarity.
  }
  \label{fig:diff_lx}
\end{figure}

We compare the XLF of clusters in superclusters with 
that of the field clusters by calculating differential and cumulative 
XLFs in Fig.~\ref{fig:diff_lx}. We consider two samples, the flux-limited 
\rtwo\ sample and the VLS reference sample. Note that the two luminosity 
functions are normalised in a different way. For the flux-limited sample
in the bottom panel of Fig.~\ref{fig:diff_lx} all three luminosity functions
(all in black, superclusters in dotted, and field in dashed line) are 
normalised by the total survey volume, $V_{\mathrm X}(L_{\mathrm X})$. Thus in 
this plot the superclusters and field luminosity functions add up to give 
the total function.

For the VLS the normalisation volume is constant and does not depend on $L_{\mathrm X}$. Here we normalised the luminosity functions by the total VLS volume,
by the volume occupied by superclusters as calculated in Sec. 6.1, and by
the residual volume of VLS after subtraction of the supercluster volume for
the total, the superclusters and the field XLF, respectively. Now the XLF
of the superclusters has a normalisation about 2 orders of magnitude higher
than the total function as the volume occupied by superclusters is very small
while more than half of the clusters reside in superclusters. The cluster
density in the field is thus lower by about a factor of 2 and so is the
normalisation of the field clusters XLF.

At first glance we do not see a striking difference in the shape of the XLF.
We therefore resort to a more subtle method for the comparison of the shape
of the function, using a Kolmogorov-Smirnov test. We restrict this test to
the VLS for the following reason. For the total \reflex\ sample the XLF
as shown in Fig.~\ref{fig:diff_lx} is affected by a bias which can distort 
our results. A close inspection of Fig.~\ref{fig:count_z} shows 
that the fraction of clusters in superclusters are about 60\% in the lower 
redshift and decreases to as low as 14\% at higher redshift. Since the high
redshift region is only populated by very luminous clusters, the field
population gets a larger share of this very luminous clusters while the
superclusters get a correspondingly larger share of the less luminous nearby
systems. We could in principle correct for this by normalising the luminosity
function by the actual volume occupied by the superclusters. However, since the
nature of the superclusters and possibly also the nature of the estimated
volume changes with redshift the introduction of another bias effect can 
hardly be avoided. 

\begin{figure}
  \centering
  \resizebox{\hsize}{!}{\includegraphics{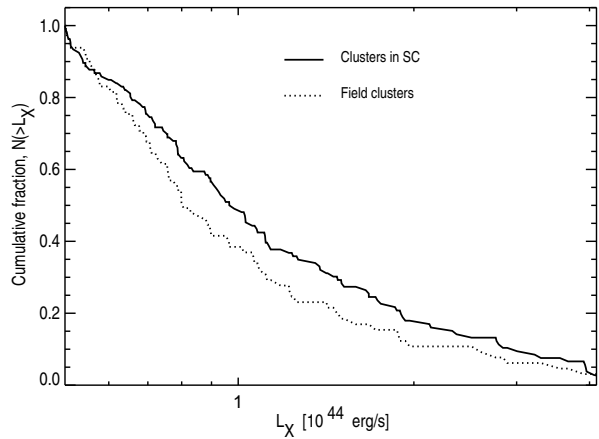}}
  \caption{
    Cumulative fractional number counts as a function of luminosity for the 
    reference volume-limited \rtwo\ cluster sample. The clusters in 
    superclusters are shown as solid line, and field clusters as dotted line. 
    The KS test of these two distributions reveals that the KS probability, 
    $P$=0.03 with a maximum separation of 0.15 at $L_{\mathrm X}=9\times10^{43}$ erg/s.
  }
  \label{fig:cum_lx}
\end{figure}

We do not have this problem for the VLS, thus we compare the normalised 
cumulative XLF of clusters in superclusters, and in the field in 
Fig.~\ref{fig:cum_lx}. Now we see a clear effect that we have a slightly more 
top-heavy luminosity function in the superclusters. 
As the distribution functions of the cluster luminosities are
un-binned, a Kolmogorov-Smirnov (KS) test was chosen to see  
if the luminosity distributions of field clusters and supercluster 
members are different. The test generally works quite well as 
long as the effective number of clusters 
$N_\mathrm{e}=\frac{N_{SC}N_{Field}}{N_{SC}+N_{Field}}\geq4$. 
The KS probability, $P$, lies between 0 and 1, and a small value 
means that the two distributions are significantly different.
The KS test of the VLS yields $P$=0.03 with a maximum separation
of 0.15 at $L_{\mathrm X}=9\times10^{43}$ erg/s. 
To assess the statistical significance of this seemingly low value of $P$ 
in an independent way we perform a null hypothesis test with simulations. We randomly re-assign 
the observed luminosities of the VLS sample to clusters, and repeat the KS 
tests on two cumulative luminosity functions 1000 times. We find that 13\% of 
the KS probabilities is lower than $P$=0.03 for this null test, which 
corresponds roughly to 1.5$\sigma$ detection of a difference in the
luminosity functions. We prefer to quote the latter value of the more
conservative probability of the null test as the significance of our results. 
Hence we conclude that although the difference of 
the luminosity distributions of the two populations of clusters is not
very large, we find an inclination for the difference using the KS test, 
and there are more luminous clusters in superclusters than in the field
in this volume-limited sample.

We therefore find the expected over-abundance of X-ray luminous clusters in 
superclusters, but the effect is relatively mild. Apart from the cosmological
explanation for this effect given above, we can think also of another possible
explanation. In dense environments we may expect a larger frequency of cluster
mergers, e.g.~\cite{richstone-92}. 
\citet*{randall-02} analysed in hydrodynamical simulations the change of the 
XLF and the temperature function of merging clusters, and found that
the X-ray luminosity is boosted towards higher values for about one
sound crossing time during a merger event. The cumulative effect of
these merger boosts affects the XLF causing an increase of the apparent number 
of hot luminous clusters. This boost effect has not been verified by 
observations, however. An important observation in this context was made 
by~\citet{schuecker-01}, which showed that the fraction of sub-structured 
clusters increases with the cluster space density, analogous to the morphology 
segregation in clusters. Substructured clusters are the products of merging 
smaller clusters. Thus superclusters as regions of cluster overdensities are 
regions of enhanced interactions of clusters. Hence our finding that the XLF 
of clusters in superclusters differs from that of field clusters 
provides some evidence that the environment where the superclusters grow is 
different from the rest of our Universe. We plan to study these questions in 
more detail in subsequent work where we compare our observation with 
cosmological simulations.

\section{Summary and outlook}

We presented the first X-ray flux-limited supercluster catalogue based on the 
\rtwo\ clusters. Using a friends-of-friends algorithm we explored 
the selection of superclusters with different overdensity parameters, and 
studied the properties of the supercluster catalogues. Our reference sample
is constructed with $f=10$, which corresponds to finding marginally 
bound objects given our cosmology. The resulting \rtwo\ supercluster 
catalogue 
comprises 164 superclusters in the redshift range $z\leq0.4$. The robustness 
of selection for varying linking lengths was tested by comparing to other 
previously known catalogues. Among the \rtwo\ superclusters are the 
well-known nearby superclusters like Shapley and Hydra--Centaurus as well as  
unknown superclusters. On average these newly found superclusters have 
higher redshifts than what is already known with cluster of galaxies thanks to 
the redshift coverage of the \rtwo\. We compared the supercluster catalogue
built with $f=10$ to that with $f=50$, which should trace the superclusters 
which correspond to the bound objects that are at turn-around. All superclusters
at $f=50$ were found in the $f=10$ catalogue.

We studied the multiplicity functions which show that our supercluster 
catalogue is largely dominated by superclusters with 5 or less cluster members
from the \rtwo\ cluster catalogue. Since superclusters are not collapsed 
objects, the definition of their volume is not unique and we decided to define 
the size of the superclusters by half the maximum extent of the clusters in 
supercluster. The distribution of this maximum radius shows that most of the 
superclusters are of size between 5 and 20 Mpc, and larger superclusters are 
dominantly found at higher redshifts. 

We constructed a volume-limited sample of the \rtwo\ catalogue 
with the constraints $z\leq0.1$ and $L_{\mathrm X}\geq$ 
5$\times10^{-43}$ erg/s.  
This control reference sample allows us to study the effect of a 
volume-limited in comparison to the flux-limited supercluster catalogue. One of our findings is 
that increasing the luminosity cut to make a different VLS emulates the effect 
of redshift on the selection of superclusters in the flux-limited sample. 
Thus, we were able to diagnose without external data a possible contamination 
in the catalogue, which turns out to be a maximum statistical contamination of
5.8\% in our catalogue for $z\leq0.22$.

Models of structure formation imply that the supercluster environment evolves 
differently from the rest of the Universe. This should result in a top-heavy 
cluster X-ray luminosity function. Thus we studied the XLF in superclusters 
and the field to test this expectation. 
Our finding suggests that in our volume-limited sample there are 
more luminous clusters in superclusters, and that the volume where 
superclusters are found is as minuscule as less than 2\% in the volume-limited 
sample. Hence we conclude that superclusters provide a good astrophysical 
ground to study the evolution of the structure growth in a distinctly dense
environment. In a future publication we perform a detailed comparison to 
N-body simulations to better understand the properties of the X-ray 
superclusters and the underlying dark matter distribution.

\section*{Acknowledgments}

GC acknowledges the support from Deutsches Zentrum f\"ur Luft und Raumfahrt 
(DLR) with the program ID 50 R 1004. We acknowledge support 
from the DfG Transregio Program TR33 and the Munich Excellence Cluster 
``Structure and Evolution of the Universe''. 
NN and HB acknowledge fruitful discussions with Peter Schuecker at 
an early stage of this work.

\bibliographystyle{mn2e}
\bibliography{paper}

\appendix

\section{Distribution of \reflex\ II superclusters}

\begin{figure}%[ht!]
  \begin{minipage}[b]{\linewidth}
    \resizebox{\hsize}{!}{\includegraphics{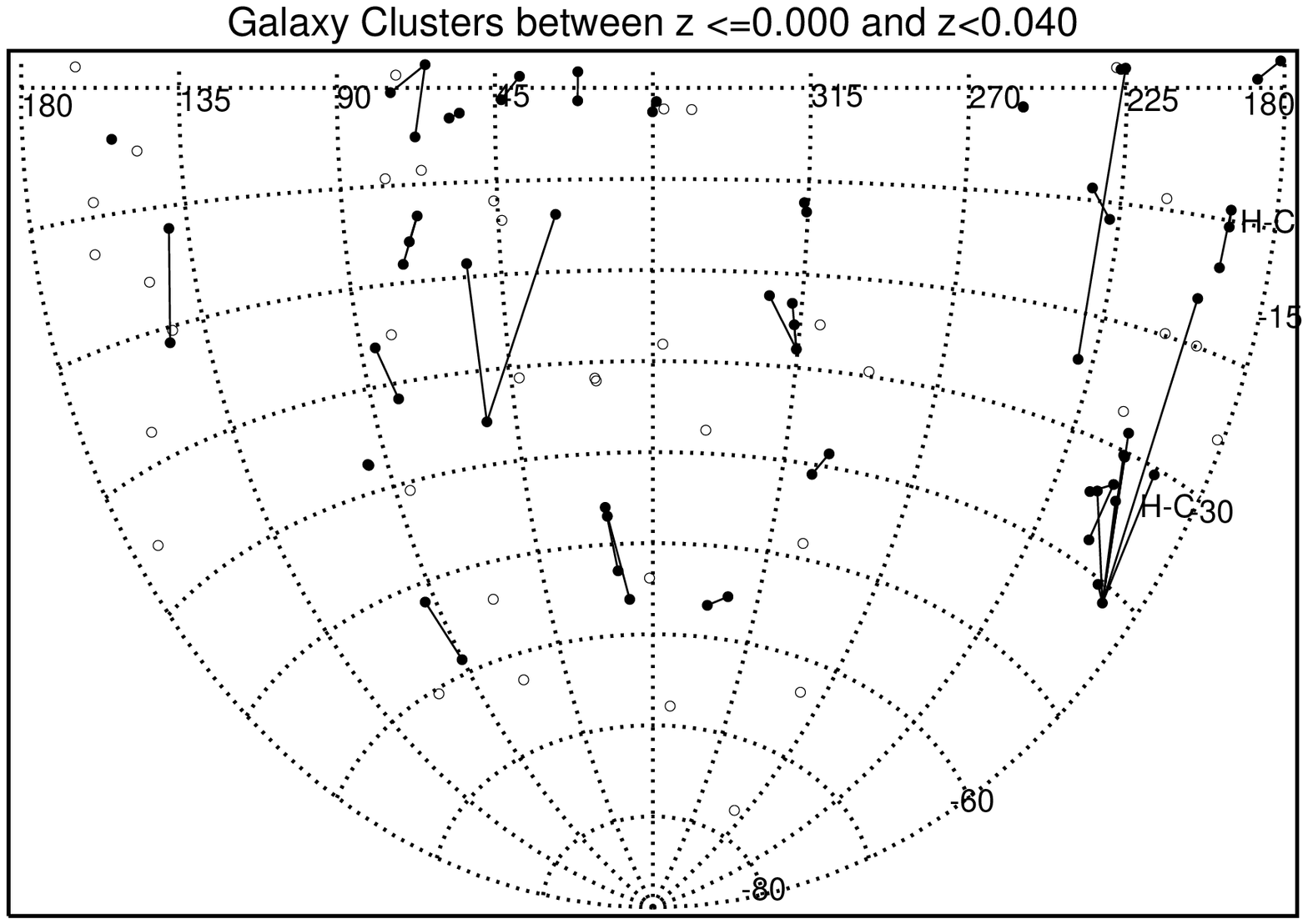}}
    \resizebox{\hsize}{!}{\includegraphics{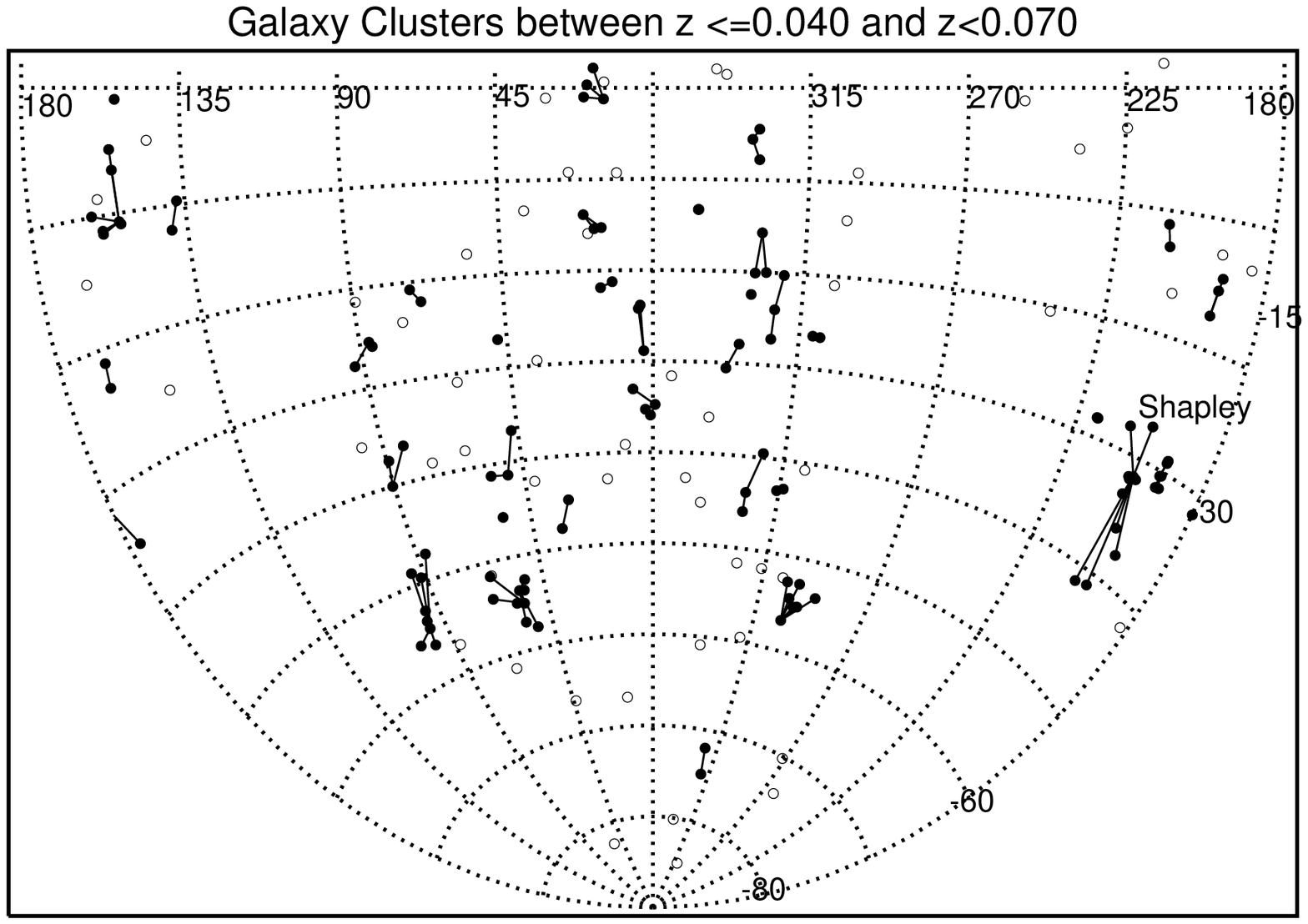}}
    \resizebox{\hsize}{!}{\includegraphics{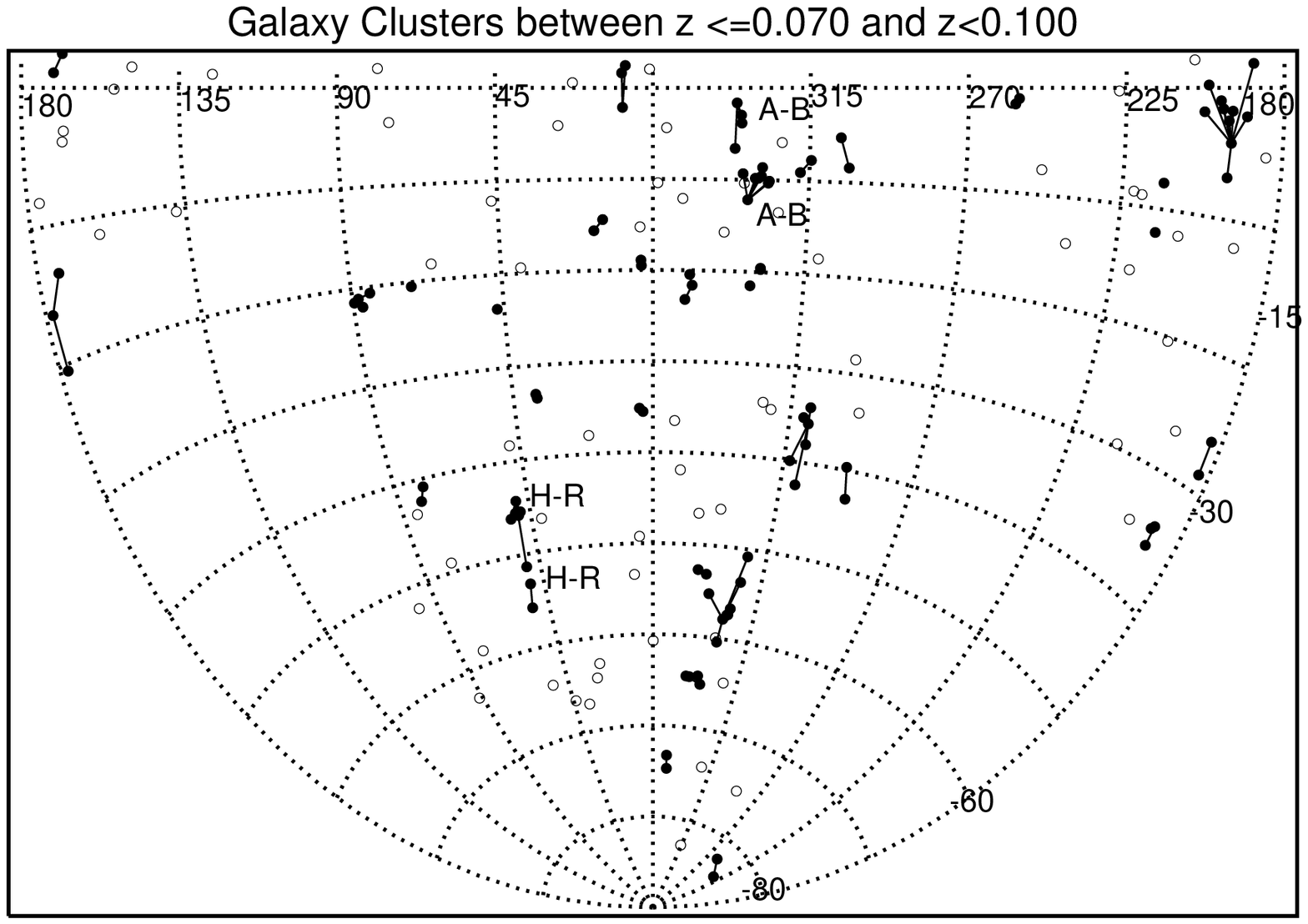}}
    \resizebox{\hsize}{!}{\includegraphics{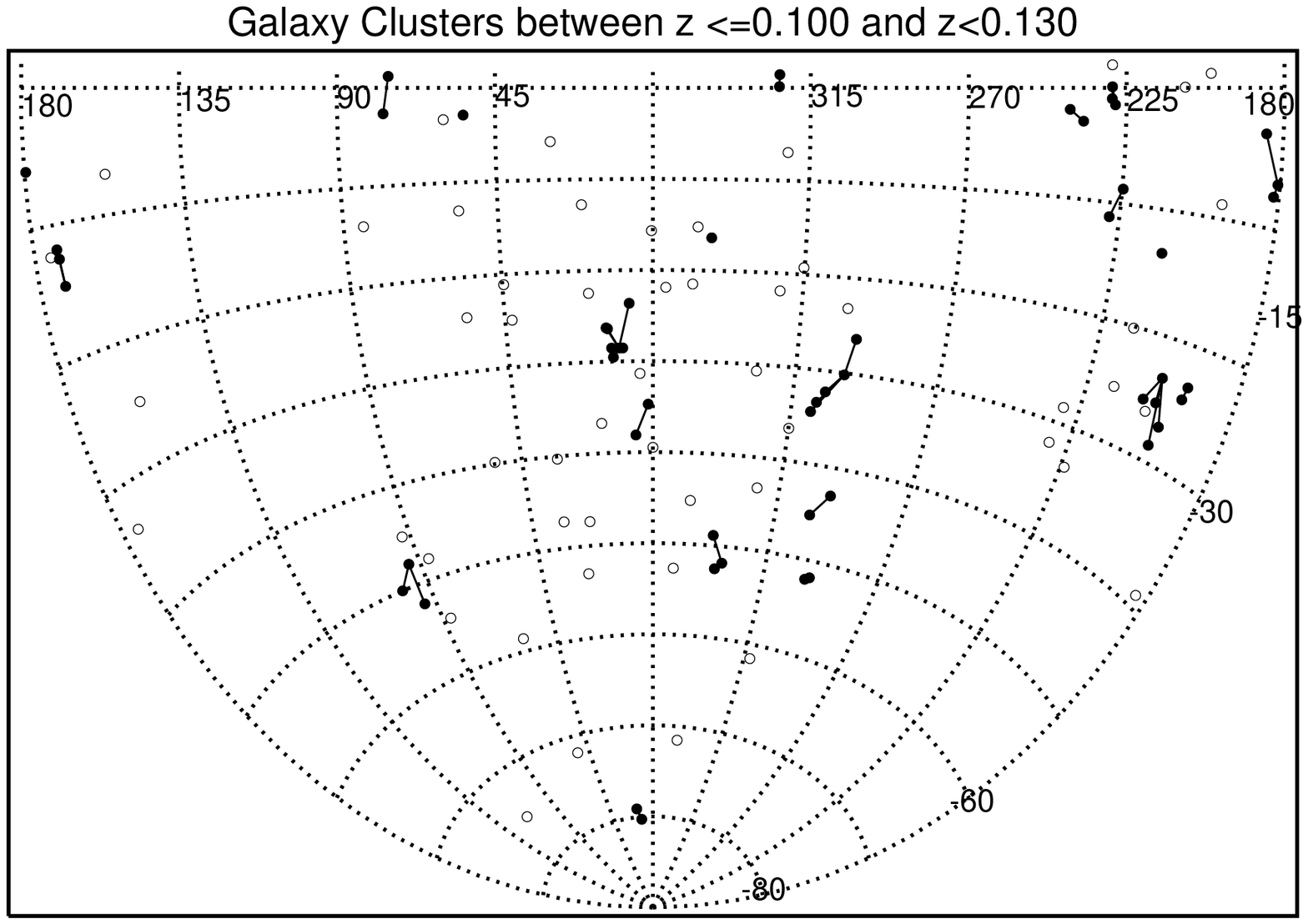}}
  \end{minipage}
\end{figure}
\begin{figure}
  \begin{minipage}[b]{\linewidth}
    \resizebox{\hsize}{!}{\includegraphics{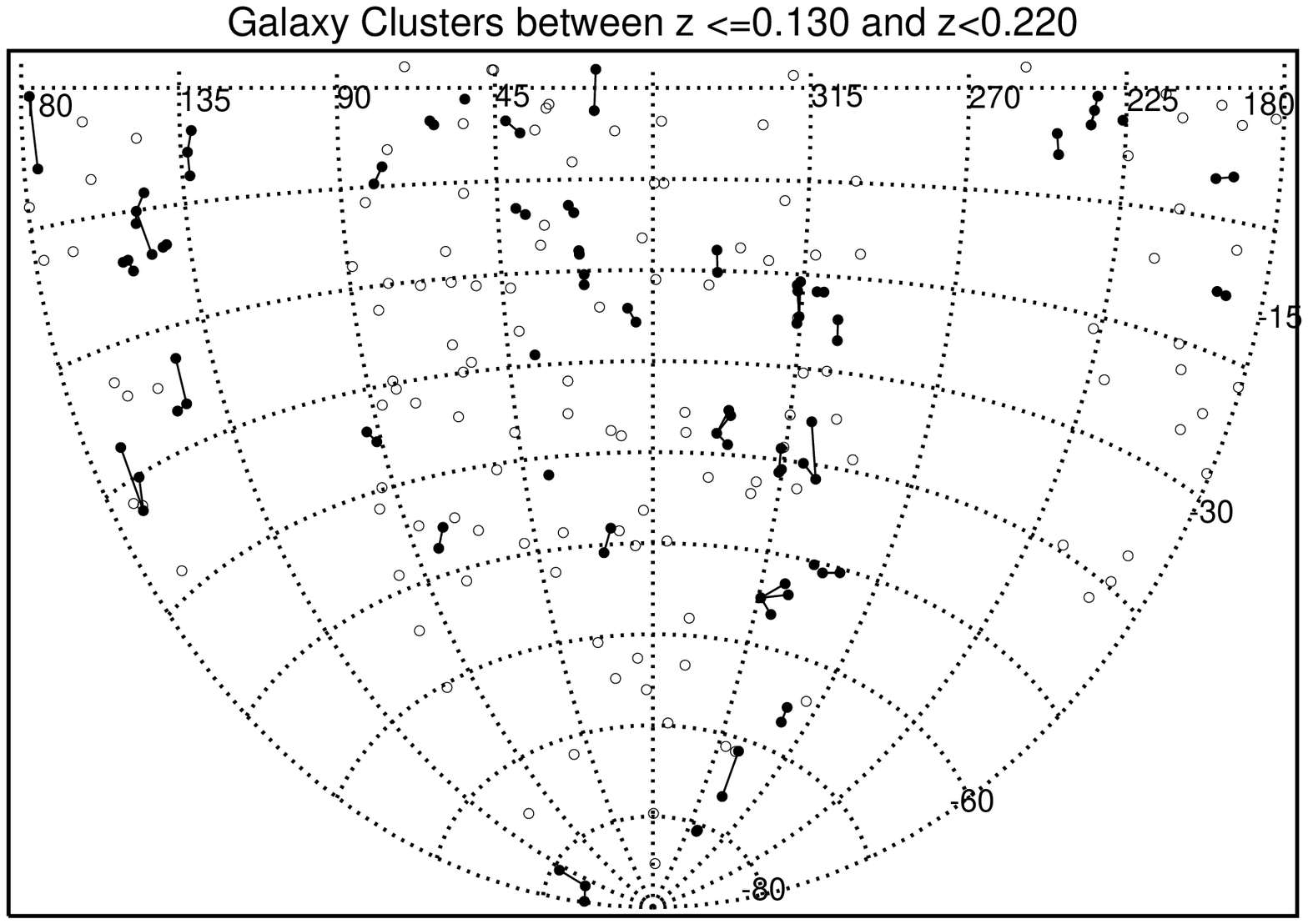}}
    \resizebox{\hsize}{!}{\includegraphics{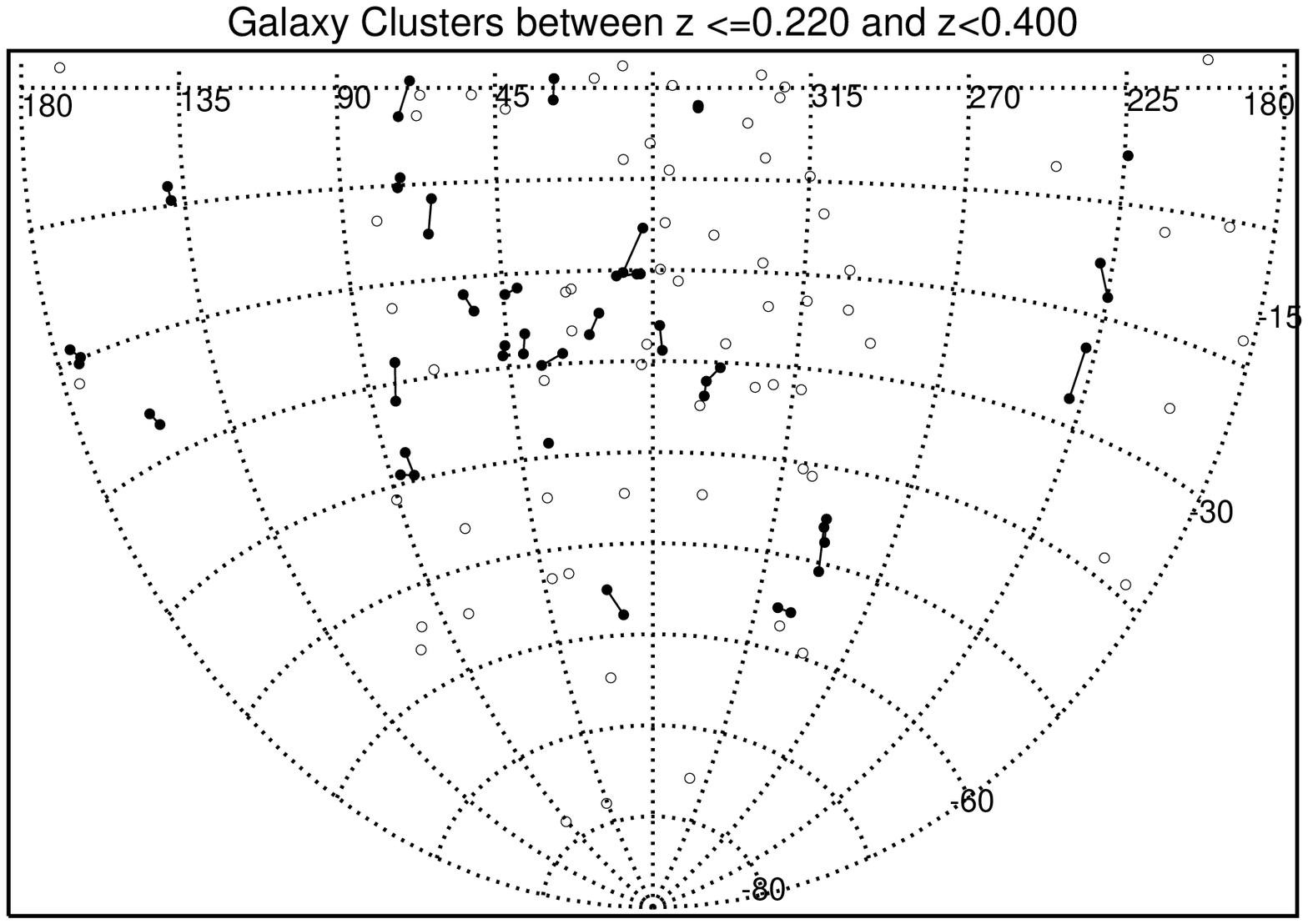}}
  \end{minipage}
  \caption{
    The spatial distribution of the \rtwo\ clusters divided
    into six redshift shells out to $z=0.4$.  
    The open circles represent field clusters and the filled circles 
    are the clusters in the superclusters.
    The supercluster members are connected to its main
    BSC by a line. We mark some of the known 
    superclusters with the abbreviations. H-C denotes Hydra-Centaurus, 
    A-B Aquarius-B, and H-R Horologium-Reticulum. 
  }
  \label{fig:scl-dist}
\end{figure}

\section{List of \reflex\ II superclusters}

\begin{multicols}{2}
\end{multicols}
\onecolumn
\begin{longtable}{c c c c c c c c c}
\caption{List of the \rtwo\ superclusters. The columns are
(1) Supercluster name
(2) R.A. (deg) (3) Dec. (deg) (4) redshift (5) $R_{\mathrm max}$  (Mpc)
(6) Total luminosity (10$^{44}$ erg/s) (7) Multiplicity 
(8) The existence in $f=50$ catalogue 
(9) ID number. The positions and redshifts are calculated
with an X-ray $L_X$ weighting of member clusters as described in the text. 
RA and Dec are for J2000. 
$R_{\mathrm max}$ is half the maximum extent between clusters in the supercluster.
$L_{\mathrm X}$ is the sum of cluster luminosities. 
The multiplicity is the number of member clusters. Column 8 indicates
if the supercluster found in the $f=10$ catalogue was found
in the $f=50$ catalogue. $Y-x$ indicates that $x$ number of clusters that 
were missing in the $f=50$ catalogue, and $Y/x$ indicates that
the supercluster was divided into $x$ number of superclusters.
The catalogue constructed with $f=50$ contains no new superclusters 
compared to the main catalogue.
}\\
\hline
\hline
Superclusters & $\alpha$[$^\circ$] & $\delta$[$^\circ$] 
& $z_\mathrm{SC}$ & $R_\mathrm{max}$ [Mpc] &  $L_X$ [10$^{44}$ erg/s] & Multiplicity
&$f$=50 & ID\\
(1) & (2) & (3) & (4) & (5) & (6) & (7) & (8) & (9) \\
\hline
\endfirsthead
\caption{continued.}\\
\hline
\hline
\endhead
\hline
\endfoot
RXSCJ2358-0203 &   359.546 &    -2.061 &   0.0380 &     2.26 &      0.137 &     2 &Y       &	 1 \\ 
RXSCJ0006-3442 &     1.561 &   -34.701 &   0.0484 &    11.48 &      2.760 &     4 &Y       &	 2 \\ 
RXSCJ0011-3557 &     2.978 &   -35.957 &   0.1163 &    22.22 &      2.558 &     2 &        &	 3 \\ 
RXSCJ0013-2639 &     3.405 &   -26.663 &   0.0629 &    12.95 &      1.638 &     3 &Y-1     &	 4 \\ 
RXSCJ0022-1745 &     5.730 &   -17.752 &   0.3729 &    99.50 &     21.060 &     2 &        &	 5 \\ 
RXSCJ0015-3518 &     3.933 &   -35.312 &   0.0960 &     5.42 &      1.076 &     2 &Y       &	 6 \\ 
RXSCJ0013-1915 &     3.454 &   -19.262 &   0.0943 &     2.66 &      1.922 &     2 &Y       &	 7 \\ 
RXSCJ0026-2031 &     6.641 &   -20.529 &   0.2943 &    81.64 &     18.453 &     3 &        &	 8 \\ 
RXSCJ0024-2512 &     6.012 &   -25.204 &   0.1400 &    14.91 &      4.033 &     2 &Y       &	 9 \\ 
RXSCJ0044-2727 &    11.142 &   -27.461 &   0.1110 &    30.93 &      8.496 &     7 &Y-1     &	10 \\ 
RXSCJ0034-0014 &     8.543 &    -0.244 &   0.0804 &    13.93 &      1.877 &     3 &Y-1     &	11 \\ 
RXSCJ0100-5004 &    15.056 &   -50.071 &   0.0249 &    13.50 &      0.356 &     4 &Y-1     &	12 \\ 
RXSCJ0059-2143 &    14.787 &   -21.723 &   0.0579 &    11.03 &      1.738 &     2 &        &	13 \\ 
RXSCJ0107-5612 &    16.819 &   -56.204 &   0.2531 &    41.39 &      8.508 &     2 &        &	14 \\ 
RXSCJ0058-7954 &    14.534 &   -79.908 &   0.1153 &    10.16 &      4.369 &     2 &Y       &	15 \\ 
RXSCJ0105-0019 &    16.341 &    -0.317 &   0.0445 &     9.91 &      2.276 &     4 &Y-1     &	16 \\ 
RXSCJ0104-1508 &    16.144 &   -15.135 &   0.0966 &    10.07 &      1.326 &     2 &Y       &	17 \\ 
RXSCJ0112-1441 &    18.224 &   -14.687 &   0.0526 &    10.48 &      1.464 &     3 &Y       &	18 \\ 
RXSCJ0110-4932 &    17.549 &   -49.534 &   0.2052 &    26.75 &      5.310 &     2 &        &	19 \\ 
RXSCJ0105-0016 &    16.495 &    -0.276 &   0.1933 &    32.76 &      5.322 &     2 &        &	20 \\ 
RXSCJ0111-2548 &    17.928 &   -25.803 &   0.2288 &    30.02 &     10.588 &     2 &        &	21 \\ 
RXSCJ0121-2055 &    20.495 &   -20.926 &   0.1720 &    11.11 &      2.939 &     2 &Y       &	22 \\ 
RXSCJ0125-0013 &    21.403 &    -0.217 &   0.0178 &     2.45 &      0.126 &     2 &Y       &	23 \\ 
RXSCJ0126-1758 &    21.748 &   -17.970 &   0.1453 &     3.40 &      2.200 &     2 &Y       &	24 \\ 
RXSCJ0134-1318 &    23.538 &   -13.308 &   0.2069 &    13.23 &      8.393 &     2 &Y       &	25 \\ 
RXSCJ0205-2929 &    31.441 &   -29.492 &   0.3888 &    80.15 &     19.491 &     2 &Y       &	26 \\ 
RXSCJ0153-0005 &    28.275 &    -0.100 &   0.2364 &    32.55 &      8.491 &     2 &        &	27 \\ 
RXSCJ0321-2857 &    50.390 &   -28.963 &   0.0052 &     5.98 &      0.025 &     3 &Y-1     &	28 \\ 
RXSCJ0210-4628 &    32.503 &   -46.470 &   0.0644 &     9.44 &      0.595 &     2 &        &	29 \\ 
RXSCJ0223-4018 &    35.884 &   -40.316 &   0.2234 &    32.53 &     10.907 &     2 &        &	30 \\ 
RXSCJ0236-2755 &    39.069 &   -27.929 &   0.2237 &    47.16 &     10.353 &     3 &        &	31 \\ 
RXSCJ0234-1317 &    38.520 &   -13.291 &   0.1644 &    16.00 &      2.822 &     2 &Y       &	32 \\ 
RXSCJ0229-3320 &    37.494 &   -33.343 &   0.0777 &     6.77 &      0.998 &     2 &Y       &	33 \\ 
RXSCJ0241-0405 &    40.338 &    -4.084 &   0.1866 &    29.46 &      6.699 &     2 &        &	34 \\ 
RXSCJ0246-0030 &    41.607 &    -0.503 &   0.0231 &     5.73 &      0.149 &     2 &Y       &	35 \\ 
RXSCJ0248-2148 &    42.221 &   -21.805 &   0.3175 &    42.03 &     19.670 &     2 &Y       &	36 \\ 
RXSCJ0305-2811 &    46.341 &   -28.197 &   0.2577 &    21.41 &     10.675 &     2 &Y       &	37 \\ 
RXSCJ0309-2519 &    47.470 &   -25.327 &   0.0696 &     9.72 &      0.655 &     2 &        &	38 \\ 
RXSCJ0326-4012 &    51.640 &   -40.201 &   0.0633 &    23.09 &      1.383 &     3 &        &	39 \\ 
RXSCJ0321-4528 &    50.273 &   -45.471 &   0.0731 &    21.62 &      6.624 &     7 &Y-1     &	40 \\ 
RXSCJ0326-5455 &    51.637 &   -54.929 &   0.0835 &    14.13 &      1.987 &     2 &        &	41 \\ 
RXSCJ0338-5414 &    54.501 &   -54.245 &   0.0603 &    19.61 &      4.901 &     5 &Y-2     &	42 \\ 
RXSCJ0335-0157 &    53.808 &    -1.957 &   0.1340 &    22.75 &      2.169 &     2 &        &	43 \\ 
RXSCJ0342-2241 &    55.598 &   -22.687 &   0.2395 &    26.40 &      8.098 &     2 &Y       &	44 \\ 
RXSCJ0403-5408 &    60.923 &   -54.150 &   0.0436 &    17.68 &      0.651 &     4 &Y/2     &	45 \\ 
RXSCJ0344-0249 &    56.076 &    -2.826 &   0.0355 &     4.55 &      0.285 &     2 &Y       &	46 \\ 
RXSCJ0412-0338 &    63.067 &    -3.648 &   0.1381 &    13.93 &      2.897 &     2 &Y       &	47 \\ 
RXSCJ0419-1328 &    64.792 &   -13.481 &   0.2962 &    71.90 &     10.644 &     2 &        &	48 \\ 
RXSCJ0428+0039 &    67.223 &     0.652 &   0.0138 &     6.84 &      0.219 &     3 &Y-1     &	49 \\ 
RXSCJ0438-1429 &    69.743 &   -14.493 &   0.0334 &     8.83 &      2.317 &     3 &        &	50 \\ 
RXSCJ0441-0035 &    70.474 &    -0.593 &   0.2784 &    55.82 &     12.015 &     2 &        &	51 \\ 
RXSCJ0445-2106 &    71.324 &   -21.105 &   0.0689 &    16.47 &      1.885 &     3 &        &	52 \\ 
RXSCJ0502-0916 &    75.719 &    -9.282 &   0.2284 &    82.43 &     22.230 &     4 &        &	53 \\ 
RXSCJ0504-0032 &    76.080 &    -0.537 &   0.1245 &    18.55 &      2.870 &     2 &        &	54 \\ 
RXSCJ0516-4610 &    79.222 &   -46.176 &   0.1961 &    28.67 &      8.746 &     2 &Y       &	55 \\ 
RXSCJ0526-3009 &    81.556 &   -30.159 &   0.3310 &    64.83 &     14.020 &     2 &        &	56 \\ 
RXSCJ0524-4124 &    81.165 &   -41.416 &   0.0758 &     5.64 &      1.014 &     2 &Y       &	57 \\ 
RXSCJ0544-2722 &    86.222 &   -27.371 &   0.0408 &    21.79 &      1.463 &     5 &Y-2     &	58 \\ 
RXSCJ0535-3842 &    83.957 &   -38.705 &   0.2906 &    95.54 &     25.664 &     3 &Y-1     &	59 \\ 
RXSCJ0554-3837 &    88.629 &   -38.633 &   0.0463 &    11.92 &      1.540 &     3 &Y-1     &	60 \\ 
RXSCJ0549-2111 &    87.402 &   -21.192 &   0.0958 &    21.88 &      4.739 &     4 &Y/2     &	61 \\ 
RXSCJ0624-5319 &    96.015 &   -53.328 &   0.0520 &    26.52 &      4.243 &     8 &Y-3     &	62 \\ 
RXSCJ0607-5559 &    91.848 &   -55.985 &   0.0362 &    10.07 &      0.157 &     2 &        &	63 \\ 
RXSCJ0609-3454 &    92.256 &   -34.914 &   0.1382 &    12.71 &      5.770 &     2 &Y       &	64 \\ 
RXSCJ0623-4937 &    95.974 &   -49.617 &   0.1194 &    23.78 &      2.806 &     3 &        &	65 \\ 
RXSCJ0625-3717 &    96.256 &   -37.293 &   0.0344 &     6.12 &      0.205 &     2 &Y       &	66 \\ 
RXSCJ0850-0535 &   132.536 &    -5.595 &   0.1842 &    36.14 &      5.557 &     3 &Y-1     &	67 \\ 
RXSCJ0913-1044 &   138.279 &   -10.749 &   0.0541 &     6.57 &      6.743 &     2 &Y       &	68 \\ 
RXSCJ0916-0855 &   139.161 &    -8.929 &   0.2256 &    18.83 &      5.183 &     2 &Y       &	69 \\ 
RXSCJ0933-1704 &   143.417 &   -17.077 &   0.0070 &     4.10 &      0.008 &     2 &Y       &	70 \\ 
RXSCJ1002-8242 &   150.543 &   -82.701 &   0.1994 &    56.54 &      7.728 &     3 &        &	71 \\ 
RXSCJ0929-1318 &   142.263 &   -13.300 &   0.1405 &    10.58 &      2.290 &     2 &Y       &	72 \\ 
RXSCJ0947-2536 &   146.800 &   -25.614 &   0.1413 &    32.69 &      5.646 &     3 &Y-1     &	73 \\ 
RXSCJ0952-1049 &   148.198 &   -10.822 &   0.1650 &    56.87 &     16.271 &     4 &Y-2     &	74 \\ 
RXSCJ1014-1405 &   153.643 &   -14.096 &   0.1499 &    21.16 &      4.750 &     3 &Y       &	75 \\ 
RXSCJ1014-0141 &   153.677 &    -1.685 &   0.0433 &    11.17 &      0.493 &     2 &        &	76 \\ 
RXSCJ1025-0934 &   156.251 &    -9.572 &   0.0554 &    19.69 &      2.053 &     7 &Y/3     &	77 \\ 
RXSCJ1026-2652 &   156.624 &   -26.877 &   0.2518 &    22.58 &     17.351 &     2 &Y       &	78 \\ 
RXSCJ1106-2210 &   166.628 &   -22.170 &   0.0640 &     4.66 &      0.711 &     2 &Y       &	79 \\ 
RXSCJ1118-3023 &   169.577 &   -30.399 &   0.1965 &    42.78 &      7.384 &     3 &        &	80 \\ 
RXSCJ1116+0206 &   169.057 &     2.110 &   0.0749 &     8.34 &      1.349 &     2 &Y       &	81 \\ 
RXSCJ1131-1334 &   172.808 &   -13.569 &   0.1048 &    16.76 &      3.841 &     3 &Y       &	82 \\ 
RXSCJ1135-1950 &   173.859 &   -19.848 &   0.3065 &    25.97 &     24.512 &     3 &Y       &	83 \\ 
RXSCJ1145-1620 &   176.419 &   -16.345 &   0.0731 &    22.98 &      1.483 &     3 &        &	84 \\ 
RXSCJ1200-0456 &   180.065 &    -4.939 &   0.1289 &    49.62 &      6.622 &     6 &Y-4     &	85 \\ 
RXSCJ1152-3259 &   178.138 &   -32.987 &   0.0691 &    11.28 &      1.497 &     2 &        &	86 \\ 
RXSCJ1210+0137 &   182.595 &     1.618 &   0.0205 &     8.17 &      0.195 &     2 &        &	87 \\ 
RXSCJ1212-2726 &   183.225 &   -27.444 &   0.0822 &     8.10 &      0.769 &     2 &Y       &	88 \\ 
RXSCJ1305-0221 &   196.451 &    -2.353 &   0.0848 &    45.01 &     14.282 &    10 &Y/2-3   &	 89 \\ 
RXSCJ1237-3429 &   189.382 &   -34.490 &   0.0763 &    13.21 &      1.697 &     3 &Y-1     &	90 \\ 
RXSCJ1250-1529 &   192.592 &   -15.493 &   0.1454 &    28.35 &      3.300 &     2 &        &	91 \\ 
RXSCJ1316-3439 &   199.020 &   -34.658 &   0.0120 &    18.57 &      1.079 &     9 &Y-2     &	92 \\ 
RXSCJ1255-1638 &   193.901 &   -16.650 &   0.0467 &     9.12 &      2.364 &     3 &Y       &	93 \\ 
RXSCJ1255-3012 &   193.808 &   -30.204 &   0.0546 &     6.24 &      5.247 &     6 &Y       &	94 \\ 
RXSCJ1254-1045 &   193.530 &   -10.763 &   0.0151 &     2.86 &      0.087 &     3 &Y       &	95 \\ 
RXSCJ1303-0650 &   195.918 &    -6.847 &   0.1928 &    37.10 &      7.236 &     2 &        &	96 \\ 
RXSCJ1302-2353 &   195.702 &   -23.899 &   0.1284 &     6.24 &      5.036 &     2 &Y       &	97 \\ 
RXSCJ1328-3233 &   202.216 &   -32.556 &   0.0482 &    28.72 &      9.852 &    12 &Y-4     &	98 \\ 
RXSCJ1333-2526 &   203.462 &   -25.435 &   0.1225 &    35.57 &      6.696 &     5 &Y-3     &	99 \\ 
RXSCJ1349-3334 &   207.327 &   -33.577 &   0.0389 &     9.34 &      4.149 &     3 &        &   100 \\ 
RXSCJ1403-1040 &   210.965 &   -10.678 &   0.0707 &    17.74 &      1.407 &     3 &Y-1     &   101 \\ 
RXSCJ1411-1242 &   212.943 &   -12.712 &   0.0995 &    12.73 &      1.889 &     2 &        &   102 \\ 
RXSCJ1435-2821 &   218.881 &   -28.356 &   0.0676 &     6.47 &      0.743 &     2 &Y       &   103 \\ 
RXSCJ1502-0333 &   225.550 &    -3.555 &   0.2164 &    29.18 &     29.656 &     2 &        &   104 \\ 
RXSCJ1504-0954 &   226.246 &    -9.901 &   0.1059 &    17.08 &      1.824 &     2 &        &   105 \\ 
RXSCJ1505-1709 &   226.322 &   -17.155 &   0.2307 &    45.18 &     20.971 &     2 &        &   106 \\ 
RXSCJ1506-0402 &   226.712 &    -4.041 &   0.0060 &     8.32 &      0.036 &     3 &Y-1     &   107 \\ 
RXSCJ1522-1012 &   230.660 &   -10.210 &   0.0243 &     6.15 &      0.076 &     2 &        &   108 \\ 
RXSCJ1514-2503 &   228.509 &   -25.066 &   0.3256 &    68.96 &     16.131 &     2 &        &   109 \\ 
RXSCJ1515-0044 &   228.817 &    -0.739 &   0.1200 &    11.57 &      4.469 &     3 &Y       &   110 \\ 
RXSCJ1537-0220 &   234.371 &    -2.350 &   0.1516 &    18.58 &      6.861 &     3 &Y-1     &   111 \\ 
RXSCJ1614-8311 &   243.631 &   -83.194 &   0.0731 &     8.24 &      3.248 &     2 &Y       &   112 \\ 
RXSCJ1555-0234 &   238.806 &    -2.567 &   0.1013 &    14.95 &      1.505 &     2 &        &   113 \\ 
RXSCJ1616-0542 &   244.093 &    -5.708 &   0.2050 &    23.02 &     23.607 &     2 &Y       &   114 \\ 
RXSCJ1657-0148 &   254.478 &    -1.808 &   0.0314 &     0.43 &      0.258 &     2 &Y       &   115 \\ 
RXSCJ1704-0115 &   256.061 &    -1.263 &   0.0914 &     3.98 &      2.713 &     2 &Y       &   116 \\ 
RXSCJ1857-6605 &   284.375 &   -66.090 &   0.1855 &    25.54 &      5.487 &     2 &        &   117 \\ 
RXSCJ1913-5042 &   288.489 &   -50.707 &   0.1540 &    28.64 &      4.453 &     2 &        &   118 \\ 
RXSCJ1928-4155 &   292.010 &   -41.928 &   0.0767 &    11.12 &      1.250 &     2 &        &   119 \\ 
RXSCJ2002-5518 &   300.601 &   -55.306 &   0.0554 &    21.88 &      7.699 &     7 &Y/2-1   &	120 \\ 
RXSCJ1940-4931 &   295.043 &   -49.527 &   0.2269 &    50.97 &     10.785 &     3 &        &   121 \\ 
RXSCJ1941-4642 &   295.285 &   -46.711 &   0.2617 &    30.99 &     10.480 &     2 &Y       &   122 \\ 
RXSCJ1950-5156 &   297.672 &   -51.942 &   0.1072 &     4.51 &      2.177 &     2 &Y       &   123 \\ 
RXSCJ1957-4405 &   299.308 &   -44.090 &   0.1106 &    21.54 &      1.872 &     2 &        &   124 \\ 
RXSCJ1956-7336 &   299.191 &   -73.614 &   0.2185 &    40.34 &     12.460 &     2 &        &   125 \\ 
RXSCJ2011-3116 &   302.819 &   -31.279 &   0.1203 &    56.46 &      8.413 &     5 &Y-2     &   126 \\ 
RXSCJ2013-5542 &   303.330 &   -55.705 &   0.2277 &    33.32 &      7.973 &     2 &        &   127 \\ 
RXSCJ2009-3949 &   302.274 &   -39.821 &   0.0191 &     2.99 &      0.097 &     2 &Y       &   128 \\ 
RXSCJ2032-5443 &   308.052 &   -54.720 &   0.1412 &    40.89 &      7.305 &     4 &        &   129 \\ 
RXSCJ2023-3915 &   305.912 &   -39.261 &   0.1455 &    41.38 &      5.079 &     3 &        &   130 \\ 
RXSCJ2014-2500 &   303.584 &   -25.007 &   0.1583 &    27.12 &     11.718 &     2 &        &   131 \\ 
RXSCJ2015-8042 &   303.784 &   -80.714 &   0.1347 &    14.28 &      3.165 &     2 &Y       &   132 \\ 
RXSCJ2019-0655 &   304.872 &    -6.920 &   0.0808 &    12.45 &      0.862 &     2 &        &   133 \\ 
RXSCJ2037-3630 &   309.412 &   -36.516 &   0.0882 &    37.66 &      7.050 &     6 &Y-3     &   134 \\ 
RXSCJ2101-2557 &   315.404 &   -25.956 &   0.0376 &    26.24 &      0.906 &     6 &Y/2-2   &	135 \\ 
RXSCJ2039-2143 &   309.892 &   -21.730 &   0.1999 &    22.79 &      7.306 &     2 &Y       &   136 \\ 
RXSCJ2059-4313 &   314.944 &   -43.225 &   0.0494 &     4.69 &      0.377 &     2 &Y       &   137 \\ 
RXSCJ2103-0823 &   315.986 &    -8.384 &   0.0806 &    11.46 &      0.905 &     2 &        &   138 \\ 
RXSCJ2103-4030 &   315.992 &   -40.505 &   0.1624 &    36.95 &      8.784 &     3 &Y-1     &   139 \\ 
RXSCJ2102-1258 &   315.640 &   -12.975 &   0.0282 &     1.31 &      0.143 &     2 &Y       &   140 \\ 
RXSCJ2104-2249 &   316.135 &   -22.822 &   0.1866 &    42.45 &     13.041 &     5 &Y/2-1   &	141 \\ 
RXSCJ2130-2359 &   322.617 &   -23.985 &   0.0652 &    19.14 &      0.765 &     3 &        &   142 \\ 
RXSCJ2128-7340 &   322.062 &   -73.674 &   0.0575 &     8.33 &      0.688 &     2 &        &   143 \\ 
RXSCJ2143-4344 &   325.831 &   -43.740 &   0.0616 &    20.04 &      1.634 &     3 &Y-1     &   144 \\ 
RXSCJ2150-5632 &   327.703 &   -56.544 &   0.0779 &    30.95 &      8.104 &     8 &Y-4     &   145 \\ 
RXSCJ2135+0055 &   323.869 &     0.930 &   0.1219 &    13.96 &      2.474 &     2 &Y       &   146 \\ 
RXSCJ2152-1815 &   328.062 &   -18.250 &   0.0591 &    13.67 &      0.859 &     3 &Y-1     &   147 \\ 
RXSCJ2201-1020 &   330.450 &   -10.348 &   0.0813 &    27.14 &      8.276 &     8 &Y-3     &   148 \\ 
RXSCJ2206-5559 &   331.744 &   -55.985 &   0.0370 &     9.13 &      0.270 &     2 &        &   149 \\ 
RXSCJ2152-1937 &   328.083 &   -19.625 &   0.0951 &     4.13 &      2.762 &     2 &Y       &   150 \\ 
RXSCJ2201-0603 &   330.458 &    -6.052 &   0.0576 &     8.43 &      1.814 &     3 &Y       &   151 \\ 
RXSCJ2202-2202 &   330.673 &   -22.050 &   0.0708 &     8.92 &      0.978 &     2 &        &   152 \\ 
RXSCJ2228-5200 &   337.026 &   -52.002 &   0.1008 &    29.39 &      6.036 &     5 &Y/2-1   &	153 \\ 
RXSCJ2222-2938 &   335.652 &   -29.647 &   0.0589 &    11.22 &      1.196 &     2 &        &   154 \\ 
RXSCJ2223-3649 &   335.973 &   -36.819 &   0.1494 &    33.21 &     19.595 &     4 &Y-2     &   155 \\ 
RXSCJ2235-6442 &   338.959 &   -64.706 &   0.0941 &    12.21 &      5.736 &     4 &Y       &   156 \\ 
RXSCJ2221-0333 &   335.436 &    -3.565 &   0.0902 &    16.56 &      5.546 &     4 &Y-1     &   157 \\ 
RXSCJ2247-3211 &   341.830 &   -32.197 &   0.2395 &    65.03 &     14.744 &     3 &        &   158 \\ 
RXSCJ2246-1812 &   341.572 &   -18.208 &   0.1299 &    34.02 &      5.573 &     3 &Y-1     &   159 \\ 
RXSCJ2306-1319 &   346.726 &   -13.325 &   0.0670 &     4.79 &      1.150 &     2 &Y       &   160 \\ 
RXSCJ2308-0205 &   347.093 &    -2.084 &   0.2998 &    15.14 &     14.380 &     2 &Y       &   161 \\ 
RXSCJ2316-2137 &   349.030 &   -21.623 &   0.0858 &    16.87 &      3.517 &     3 &Y-1     &   162 \\ 
RXSCJ2318-7350 &   349.648 &   -73.850 &   0.0982 &     5.29 &      1.607 &     2 &Y       &   163 \\ 
RXSCJ2350-2650 &   357.683 &   -26.844 &   0.2262 &    22.75 &     15.227 &     2 &Y       &   164
\label{tab:r2cat}
\end{longtable}

\label{lastpage}

\end{document}